\documentclass[10pt,letterpaper]{article} 

\usepackage[margin=1in]{geometry}
\usepackage[colorlinks=true,allcolors=blue]{hyperref}

\usepackage{amsmath}
    
\usepackage{amssymb}    
\usepackage{bm}    

\usepackage{graphicx}   
\usepackage{subcaption}
\usepackage{multirow}
\usepackage{geometry}
\usepackage[utf8]{inputenc}

\usepackage{authblk} 

\title{Effects of Quantum Noise and Source Blurring on Dark-Field Signal Retrieval in X-ray Speckle-Based Imaging}

\author[1]{Hunwoo Lee}
\author[1]{Jingcheng Yuan}
\author[1,2,3,*]{Mini Das}

\affil[1]{Department of Physics, University of Houston, 3507 Cullen Blvd, Houston, TX 77204, USA}
\affil[2]{Department of Electrical and Computer Engineering, University of Houston, 3507 Cullen Blvd, Houston, TX 77204, USA}
\affil[3]{Department of Biomedical Engineering, University of Houston, 3507 Cullen Blvd, Houston, TX 77204, USA}
\affil[*]{Corresponding author: mdas@uh.edu}
\date{\today}

\begin{document}
\maketitle 

\begin{abstract}   

X-ray speckle-based dark-field imaging offers high sensitivity to sub-pixel structural features, yet its quantitative reliability in clinical and preclinical settings remains constrained by low photon flux and finite focal spot sizes. However, how hardware-induced noise and source blurring propagate through retrieval algorithms to degrade signal integrity is not fully understood. Here, we systematically evaluate algorithm robustness—quantified by signal linearity, sensitivity, and bias—under photon starvation and source blurring across two mathematically distinct frameworks: differential-based intrinsic tracking (Low-Coherence System, LCS) and patch-wise explicit tracking (X-ray Speckle-Tracking Speckle-Vector-Tracking, XST-XSVT).
Our experimental results demonstrate that input speckle pattern distortions propagate through retrieval algorithms in fundamentally different ways depending on algorithm architecture. As an example, using our setup, under severe photon starvation (exposure reduced from $50\text{ s}$ to $1\text{ s}$ per mask step), derivative noise amplification in LCS causes its dark-field signal linearity and sensitivity to drop precipitously by $85.7\%$ and $91.3\%$, respectively, while sharply elevating baseline bias. In contrast, XST-XSVT restricts these losses to $37.4\%$ for linearity and $64.2\%$ for sensitivity while maintaining a stable baseline, as its patch-wise variance calculation inherently suppresses stochastic noise. Similarly, under blur-limited conditions (expanding focal spot size from $7\ \mu\text{m}$ to $50\ \mu\text{m}$), source blurring washes out the speckle pattern, directly reducing dark-field sensitivity for both LCS (by $47.8\%$) and XST-XSVT (by $42.6\%$). Beyond this shared sensitivity loss, the pattern smoothing causes the differential equations in LCS to become mathematically unstable, degrading its linearity by $5.3\%$ and elevating baseline bias. Conversely, XST-XSVT robustly withstands the smoothed pattern, bypassing this instability to maintain linearity with a negligible $1.0\%$ drop.
This characterization establishes operational boundaries for low-power and low-coherence X-ray systems, guiding algorithm selection and framework optimization to realize quantitative dark-field imaging in preclinical and clinical applications.

\end{abstract}  

\section{Introduction}
X-ray imaging is a non-destructive modality for revealing the internal structures and material properties of an object\cite{als2011elements}. However, conventional absorption-based techniques suffer from poor contrast when imaging weakly attenuating materials, such as soft tissues, and lack sensitivity to sub-resolution features \cite{fitzgerald2000phase,pfeiffer2008hard}. To overcome these limitations, X-ray phase-contrast and dark-field imaging have been developed\cite{pfeiffer2008hard,endrizzi2014hard,berujon2012x,paganin2019x}. These methods exploit phase shifts and small-angle X-ray scattering, respectively. In the diagnostic energy range, the phase-shift coefficient of low-atomic-number materials is significantly larger than the absorption coefficient, particularly at lower photon energies \cite{bravin2013x,momose2020x}. Complementary to phase refraction, the dark-field signal provides a unique signature of stochastic microstructures below the system’s spatial resolution\cite{rigon2008generalized,yashiro2010origin,lynch2011interpretation,paganin2019x}.

While initially confined to synchrotrons due to stringent requirements for high brilliance and spatial coherence \cite{snigirev1995possibilities,momose1995demonstration,chapman1997diffraction}, the work by Wilkins et al and advancements in grating-based imaging and edge-illumination enabled laboratory-based implementations using polychromatic, divergent sources \cite{wilkins1996phase, pfeiffer2006phase,olivo2007coded}. These advancements paved the way for various laboratory-based approaches. 

Conventional multi-grating techniques exhibit significant limitations: the fabrication of large, small-pitch gratings remains challenging, thereby restricting the field of view \cite{astolfo2017large,bachche2017laboratory}, and these systems require precise optical alignment among multiple components \cite{pfeiffer2006phase,endrizzi2014hard}. Furthermore, they fundamentally rely on multi-exposure protocols, which inherently increase the radiation dose delivered to the specimen. To mitigate the complexities associated with grating-interferometry and edge-illumination configurations both of which require movement of X-ray optical components, single-mask approaches have been proposed to enable single-exposure implementation \cite{yuan2024transport,yuan2025single}. Although these single-mask architectures successfully eliminate the need for inter-grating alignment, they nevertheless suffer from the challenges of fabricating large-area structured masks and still require alignment between the detector pixels and the grating features.

Speckle-based X-ray imaging (SBXI) has recently emerged as a prominent technique, validated with both synchrotron and laboratory-based X-ray sources\cite{berujon2012two,morgan2012x,zanette2014speckle,wang2016synchrotron}. Its reliance on a single random mask offers significant advantages: the masks are simple and cost-effective to fabricate, the system has a larger field of view, and the alignment requirements are substantially relaxed. A typical SBXI setup, as depicted in Fig. 1 (a), comprises an X-ray source, a random mask, and an X-ray detector.  

The speckle pattern is generated as X-rays traverse the mask, arising from absorption, scattering, or a combination of both \cite{pavlov2021directional}. This study focuses specifically on absorption-based speckle patterns. This is because the conventional X-ray systems that are the subject of our work typically lack the high spatial coherence required for interference effects to occur; consequently, absorption becomes the dominant mechanism for speckle generation\cite{wang2016synchrotron}. The SBXI methodology requires the acquisition of at least one pair of images: a reference image taken with only the mask in the beam path and a sample image captured with both the mask and the sample. The introduction of the sample modulates the reference speckle pattern in three primary ways: a reduction in mean intensity due to attenuation, a transverse shift due to refraction, and localized blurring caused by small-angle scattering. A specialized retrieval algorithm is subsequently required to extract these encoded physical signals. While these algorithms are capable of retrieving all three contrast modalities simultaneously, this work focuses specifically on the extraction of the dark-field signal.

To extract the dark-field signal, an image processing step termed dark-field retrieval is performed. Various algorithms have been developed for SBXI, which are broadly classified into explicit and intrinsic speckle-tracking approaches \cite{pavlov2020x}. Explicit tracking algorithms locally analyze sample-induced speckle modulations using cross-correlation or error-minimization\cite{wang2015x,berujon2012x,berujon2015near,berujon2016x,zanette2014speckle,zdora2017x}. In contrast, intrinsic tracking approaches monitor these modulations intrinsically by solving a continuity or Fokker-Planck equation to retrieve the dark-field signal without local feature tracking \cite{pavlov2020x,alloo2023m,magnin2023dark,magnin2025x}.

While successfully adapted for laboratory settings, these techniques rely on conventional X-ray tubes where the maximum permissible power is dictated by the focal spot size. To maintain high spatial resolution, the tube must operate at a lower power to avoid thermal damage to the anode, resulting in a reduced photon flux that necessitates extended acquisition times. Furthermore, the finite focal spot size induces geometric source blurring. These inherent laboratory constraints directly compromise speckle integrity and threaten the strict linearity of the retrieved dark-field signal with respect to object thickness \cite{wang2009quantitative, vittoria2017retrieving, doherty2023edge}. Maintaining signal linearity is a prerequisite for quantitative accuracy and in particular three-dimensional computed tomography to yield material-specific parameters \cite{rigon2008generalized, wang2009quantitative, bech2010quantitative, paganin2023paraxial, doherty2023edge}. Conversely, a breakdown in this linear relationship causes severe structural distortions during volumetric reconstruction \cite{bech2010quantitative, vittoria2017retrieving, doherty2023edge}. Consequently, systematically characterizing the impact of photon noise and source blurring on this linear scaling behavior is essential to define the exact operational boundaries and limits of specific retrieval frameworks under practical laboratory conditions. 

Existing literature has largely concentrated on algorithm robustness and noise propagation in differential phase-contrast imaging, leaving the systematic characterization of the dark-field signal relatively incomplete \cite{rouge2021comparison, quenot2021implicit, celestre2025review, zdora2017x, zhou2016noise, quenot2022transfer, pavlov2020x, beltran2023fast}. However, it should be noted that several recent investigations have specifically focused on evaluating the scaling linearity of the dark-field signal. For instance, Vittoria et al. \cite{vittoria2017retrieving} observed severe sub-linear degradation using single-shot tracking configurations, a phenomenon stemming from their reliance on interference-generated speckles that undergo distortion under polychromatic X-ray illumination. In contrast, Magnin et al. \cite{magnin2025x} reported preservation of the linearity under the Low-Coherence System (LCS) framework—albeit with a non-zero background offset—by employing absorption-based speckles identical to those utilized in this work. Despite these preliminary insights, a critical research gap persists regarding the comprehensive characterization of the dark-field signal under the specific constraints imposed by conventional X-ray tubes.

To address these challenges, this study systematically evaluates how source-induced geometric blurring and photon noise degrade dark-field signal fidelity under controlled laboratory conditions. We first analyze noise- and blur-induced structural deformations in raw speckle patterns modulated by controlled variations in focal spot size and exposure time, quantifying pre-retrieval pattern dissimilarity between reference and sample acquisitions. We hypothesize that these input pattern mismatches propagate through the mathematical inversion process, driving degradations in dark-field response linearity, sensitivity, and bias. Using linear regression analysis, we map how these input distortions translate into signal degradation and demonstrate that the extent of this performance loss is strongly governed by the retrieval architecture—contrasting an unregularized, point-wise differential framework with a patch-wise variance calculation framework .

\begin{figure}[h!] 
    \centering 
        \begin{subfigure}{0.48\linewidth}
            \centering
            \includegraphics[width=\linewidth]{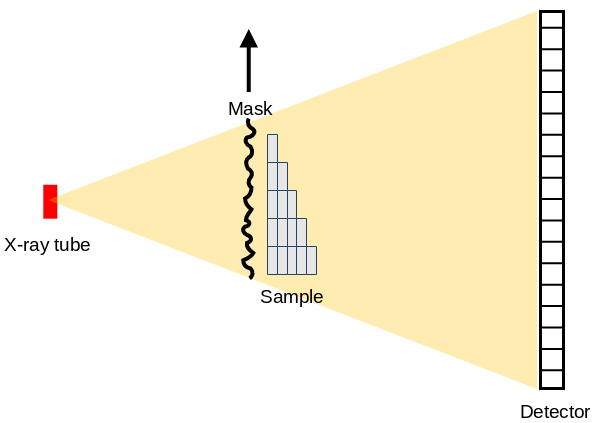} 
            \caption{} %
            \label{fig:1a_setup} 
        \end{subfigure}
        \hfill
        \begin{subfigure}{0.48\linewidth}
            \centering
            \includegraphics[width=\linewidth]{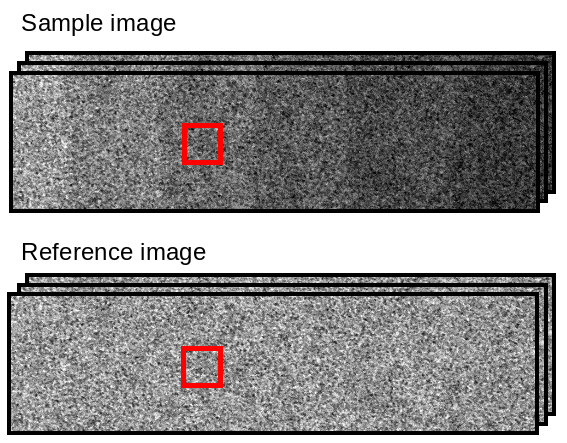} 
            \caption{} %
            \label{fig:1b_IsIr} 
        \end{subfigure}    
    \caption{(a) Schematic of the speckle-based X-ray imaging setup. (b) A representative pair of images acquired with the setup, showing a sample image and a corresponding reference (speckle-only) image. The red box indicates the analysis window used for the explicit speckle tracking method.}
    \label{fig1} 
\end{figure}


\section{Materials and methods}
\subsection{Speckle degradation due to photon noise and source blurring}
The clinical translation of SBXI relies on conventional X-ray tubes, which present significant challenges to image quality. Specifically, their inherent characteristics—finite focal spot sizes and low photon flux—introduces source blurring and photon noise. These factors directly deform the speckle pattern, which is the critical determinant of signal retrieval performance. A large focal spot blurs the speckles, reducing their visibility, while a low number of photons leads to statistical fluctuations that introduce a structural mismatch between measurements taken at different time points. Other than sample-induced effects, the speckle structures in the sample ($I_s$) and reference ($I_r$) images should be identical for accurate tracking; however, in reality, this can be compromised by temporal fluctuations and noise combined with focal spot blurring. Such degradation hinders the  accuracy of tracking algorithms and reduces the sensitivity of the retrieved dark-field signal. Furthermore, the extent of this impairment differs significantly between retrieval algorithms due to their fundamentally distinct tracking strategies (e.g., local window analysis versus global estimation from the entire image).
To quantify the speckle pattern, we employ the visibility metric ($V$), defined as the local standard deviation normalized by the mean intensity \cite{zdora2018state}:
\begin{equation} 
V = \frac{\sigma}{\bar{I}},
\end{equation}
where $\sigma$ and $\bar{I}$ represent the standard deviation and the mean of the pixel values in the window, respectively.
However, since visibility alone does not fully capture morphological changes or temporal instabilities, we also introduce two complementary image similarity metrics to compare the sample image ($I_s$) directly to the reference image ($I_r$). Specifically, we evaluate these metrics within sample-free regions where the speckle structure should otherwise remain constant. The first, the Pearson correlation coefficient assesses the preservation of the underlying structural pattern. The second, the mean squared error provides a measure of the absolute, pixel-wise dissimilarity between the two images.
The Pearson correlation coefficient evaluates the preservation of the speckle structure by measuring the linear relationship between the pixel intensity values of $I_s$ and $I_r$. The formula is given by:
\begin{equation}
\text{PCC} = \frac{\sum_{i=1}^{M} \sum_{j=1}^{N} (I_s(i, j) - \mu_s)(I_r(i, j) - \mu_r)}{\sqrt{\sum_{i=1}^{M} \sum_{j=1}^{N} (I_s(i, j) - \mu_s)^2} \sqrt{\sum_{i=1}^{M} \sum_{j=1}^{N} (I_r(i, j) - \mu_r)^2}},
\end{equation}
where $\mu_s$ and $\mu_r$ are the mean intensity values of the respective images. The Pearson correlation coefficient ranges from $-1$ to $1$, where a value of $+1$ indicates perfect structural congruency; a decrease from this ideal value quantifies the extent of the structural mismatch between the images. The mean squared error measures the average of the squared intensity differences and is calculated as:
\begin{equation}
\text{MSE} = \frac{1}{M \times N} \sum_{i=1}^{M} \sum_{j=1}^{N} [I_s(i, j) - I_r(i, j)]^2,
\end{equation}
where $M$ and $N$ are the number of pixels along the $x$ and $y$ directions, respectively. Theoretically, for an unaltered and perfectly stable speckle pattern, the Pearson correlation coefficient should be one and the mean squared error should be zero in regions unaffected by the sample. Therefore, any structural mismatch between time-separated measurements—whether from noise or blurring—will manifest as a decrease in the Pearson correlation coefficient and an increase in the mean squared error.

\subsection{Experimental method}
\begin{figure}[h!] 
    \centering 
        \begin{subfigure}{0.55\linewidth}
            \centering
            \includegraphics[width=\linewidth]{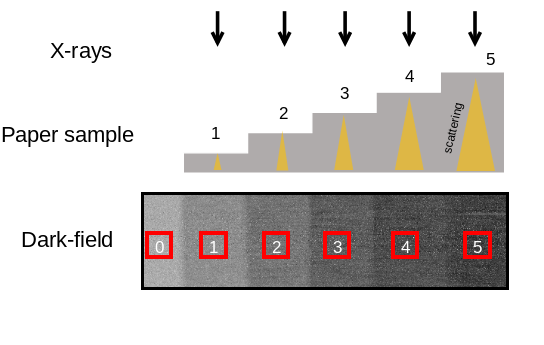} 
            \caption{} %
            \label{fig:2apaper} 
        \end{subfigure} 
        \hfill 
        \begin{subfigure}{0.4\linewidth}
            \centering
            \includegraphics[width=\linewidth]{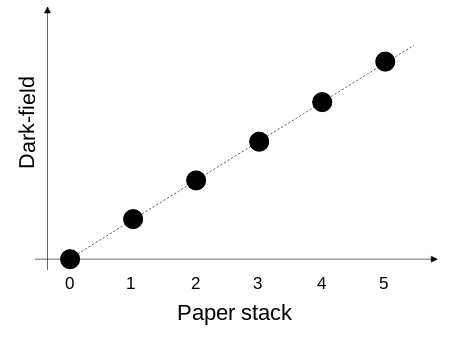} 
            \caption{} %
            \label{fig:2bplot} 
        \end{subfigure}  
    \caption{Evaluation of the linearity and sensitivity of the retrieved dark-field signal. (a) Dark-field images obtained for an increasing number of paper stacks along the beam path across its transverse profile. The six red boxes denote the regions of interest, corresponding to 0 to 5 paper stacks, from which the mean dark-field signal was calculated. (b) The corresponding plot of the mean dark-field signal as a function of paper stack thickness.}
    \label{fig2} 
\end{figure}

Speckle-based X-ray imaging relies on various speckle tracking algorithms to retrieve information about X-ray interactions with matter. Each algorithm employs a distinct approach to extract the signals corresponding to transmission, differential phase-contrast, and dark-field. As outlined in the introduction, these retrieval algorithms are broadly categorized into two families: explicit and intrinsic methods. Explicit tracking methods include Single-Shot X-ray Speckle-Tracking \cite{wang2015x}, X-ray Speckle-Scanning \cite{berujon2012x}, X-ray Speckle-Vector-Tracking \cite{berujon2015near,berujon2016x}, and Unified Modulated Pattern Analysis \cite{zanette2014speckle,zdora2017x}. Conversely, intrinsic tracking methods encompass Multimodal Intrinsic Speckle-Tracking \cite{pavlov2020x,alloo2023m} and LCS \cite{magnin2023dark,magnin2025x}.

\subsubsection{Explicit speckle tracking}

Explicit speckle tracking methods operate by analyzing and comparing local windows of pixels between a sample image ($I_s$) and a reference, speckle-only image ($I_r$), as depicted in Fig. 1 (b). These local windows can be defined by using a group of adjacent pixels, as in single-shot X-ray speckle-tracking \cite{berujon2012two,morgan2012x} or data from different mask positions at a single pixel location \cite{berujon2012x,zdora2017x}. The transmission, differential phase-contrast, and dark-field signals are then extracted by quantifying the modulation between the reference and sample windows, typically by minimizing an error function \cite{zanette2014speckle,zdora2017x} or calculating their cross-correlation \cite{berujon2012two,morgan2012x}. For this study, X-ray Speckle-Tracking X-ray Speckle-Vector Tracking(XST-XSVT) algorithm was selected as the representative explicit method. The XST-XSVT algorithm is a hybrid algorithm that enhances the conventional X-ray speckle-vector tracking method by incorporating a 1D scanning mode, which significantly reduces the number of mask steps and the total acquisition time compared to 2D scanning techniques \cite{berujon2016x}. This approach circumvents the demanding hardware requirements of pure X-ray speckle-tracking (e.g., ultra high-resolution detectors) and the high dose of 2D raster-scanning X-ray speckle-scanning, making it highly suitable for practical laboratory systems. Our choice is further supported by literature identifying both the XST-XSVT and the Unified Modulated Pattern Analysis algorithm as well-suited for laboratory-based X-ray systems \cite{celestre2025review}. Given their methodological similarities, XST-XSVT was chosen to represent this class of algorithms in our comparison \cite{rouge2021comparison}. The core of the XST-XSVT method is to find the speckle displacement by maximizing the cross-correlation coefficient between image pairs acquired at different mask positions. The algorithm involves defining a small two-dimensional window, $w$, of size $(2m+1) \times (2n+1)$ around a pixel of interest, $(x_i, y_j)$. The intensity values within this window are collected across all $K$ images with random transverse shifts to form a three-dimensional sample vector, $I_s(x, y, k)$, for all $(x, y) \in w$. This 3D sample vector is then cross-correlated with a set of 3D reference vectors, $I_r(x+\delta_x, y+\delta_y, k)$, acquired in the absence of the sample (with only the ramdom mask in the beam path) where $(\delta_x, \delta_y)$ is a displacement vector within a larger search interval, $W$. The local transverse speckle displacement, $D_{\perp}(x_i, y_j)=(\delta_x, \delta_y)$, is determined by maximizing this correlation. This process is mathematically described by the following equation\cite{celestre2025review}:
 \begin{equation}
 D_{\perp}(x_i, y_j) = \underset{(\delta_x, \delta_y) \in W}{\mathrm{argmax}} \sum_{k=1}^{K} \iint_{(x,y) \in w} S(x, y, k) \cdot R(x+\delta_x, y+\delta_y, k) \,dx\,dy,
 \end{equation}
where $S$ and $R$ represent the zero-normalized intensity vectors for the sample and reference images, respectively.
This displacement is directly related to X-ray refraction, which forms the basis of the differential phase-contrast signal. Once the displacement vector $(\delta_x, \delta_y)$ is identified, the dark-field signal is extracted by quantifying the reduction in local speckle visibility between the sample vector set, $I_s(x, y, k)$, and the aligned reference set, $I_r(x+\delta_x, y+\delta_y, k)$, across all $K$ mask positions \cite{zdora2015simulations}. Quantitatively, this signal is expressed as the ratio of sample visibility ($V_s$) to reference visibility ($V_r$) \cite{zdora2018state}:
\begin{equation}
\frac{V_s}{V_r} = \frac{\sigma_s}{\sigma_r}\frac{\bar{I_r}}{\bar{I_s}},
\end{equation}
where the subscripts $s$ and $r$ denote values derived from the sample and reference images, respectively. Thus, $\sigma_s$ and $\sigma_r$ are the local standard deviations, and $\bar{I_s}$ and $\bar{I_r}$ are the local mean intensities of the sample and reference images, respectively.

\subsubsection{Intrinsic speckle tracking}
Intrinsic speckle tracking methods are formulated based on a continuity equation—specifically, the paraxial X-ray Fokker-Planck equation combined with an optical-flow formalism \cite{paganin2019x,pavlov2020x,alloo2023m}. This approach models the transition from the reference speckle pattern ($I_r$, acquired with mask only) to the raw sample image ($I_s$, acquired with both mask and sample) as a superposition of two distinct physical processes: 1) a geometric flow describing macroscopic speckle displacement due to refraction, and 2) a diffusive flow accounting for local speckle decorrelation caused by small-angle X-ray scattering.To isolate speckle phase and scattering modulations from pure attenuation, the raw sample image $I_s(x,y)$ is normalized by the object’s intrinsic transmission map, $I_{obj}(x,y)$ (i.e., pure attenuation without speckle). As a representative intrinsic method, we selected the LCS framework developed for a low-coherence system \cite{quenot2021implicit,magnin2023dark}, which was validated for conventional laboratory X-ray sources characterized by larger focal spots and detector pixel sizes. The governing continuum relation is mathematically described by \cite{magnin2023dark}:
\begin{equation}
I_r(x,y)-\frac{I_s(x,y)}{I_{obj}(x,y)} =D_\perp(x,y) \nabla_\perp[I_r(x,y)]-z D_{intrinsic}(x,y)\nabla^2_\perp[I_r(x,y)],
\end{equation}
where $I_{obj}$, $D_\perp(x,y)$, and $D_{\text{intrinsic}}$ denote the transmission, displacement, and dark-field signals, respectively, and $z$ represents the propagation distance. These signals are simultaneously retrieved across $K$ distinct mask positions by solving the system of equations \cite{pavlov2020x}:
\begin{equation}
I^{(k)}_r(x,y)=\frac{I^{(k)}_s(x,y)}{I_{obj}(x,y)}+ \delta_x \frac{\partial I^{(k)}_r(x,y)}{\partial x}+\delta_y \frac{\partial I^{(k)}_r(x,y)}{\partial y}-z D_{intrinsic}(x,y)\nabla^2_\perp[I^{(k)}_r(x,y)].
\end{equation}
Crucially, this signal retrieval framework exhibits two interconnected structural vulnerabilities. First, applying the second-order spatial Laplacian operator directly to the reference speckle intensity acts as a high-pass filter, leaving the system highly sensitive to photon noise under photon-starved conditions. Second, spatial blur dampens the magnitude of $\nabla^2_\perp I_r$, directly diminishing intrinsic dark-field sensitivity and rendering the local coefficient matrix ill-conditioned. When the linear system is inverted on an unregularized point-wise (pixel-by-pixel) basis, this ill-conditioned matrix catastrophically multiplies the stochastic photon noise. Together, these mechanisms trigger numerical instability.


\subsection{Experimental settings}

Our imaging system consists of a polychromatic microfocus X-ray tube (Hamamatsu) operating at 40 kV with variable focal spot mode of small, medium, and large (7,20, and 50 $\mu$m), and a photon-counting detector (MediPix, Advacam) featuring a 0.5 mm thick silicon sensor and a 1280 $\times$ 256 array of 55 $\mu$m pixels. A stack of seven 150-grit sandpapers, with a nominal grain size of 111 $\mu$m, served as the random mask. The geometric configuration was established with source-to-sandpaper, sandpaper-to-sample, and sample-to-detector distances of 370, 100, and 730 $\text{mm}$, respectively. Given a system magnification of 3.24, the nominal focal spots of 7, 20, and 50 $\mu$m are projected to approximately 15, 45, and 112 $\mu$m, respectively, at the detector plane. 

For each experimental condition, ten pairs of speckle images were acquired. Each pair consisted of an image with the sample present and one without it, captured while translating the mask horizontally between acquisitions. The dark-field signals were retrieved using Eq. (5) and (7) for the explicit and intrinsic methods, respectively. In the explicit method, the signal is determined from the reduction in visibility. In contrast, the signal from the intrinsic method is proportional to the variance of the X-ray intensity point spread function\cite{paganin2023paraxial,doherty2023edge}. Theoretically, when the sample image is modeled as a convolution of the reference image and the sample's point spread function, the negative logarithm of the visibility reduction is proportional to the variance of this point spread function\cite{vittoria2017retrieving,wang2016synchrotron}. Therefore, to enable a direct comparison with the signal from the intrinsic method, the dark-field signal from the explicit method was transformed as follows\cite{vittoria2017retrieving}:
\begin{equation} 
D_{explicit}=-ln\left(\frac{V_s}{V_r}\right) \propto D_{intrinsic}.
\end{equation}

The influence of photon noise and source blurring on the speckle pattern was first characterized. Speckle visibility was calculated from reference images (Eq. (2)), while the Pearson correlation coefficient and the mean squared error were computed between image pairs. These metrics, evaluated in sample-free regions of interest, quantified system-induced signal decorrelation independent of sample effects.

The primary experiment evaluated the linearity and sensitivity of the dark-field signal using a paper phantom, with a methodology adapted from Bech et al. \cite{bech2010quantitative} and Vittoria et al. \cite{vittoria2017retrieving}. The phantom’s thickness was varied from one to five stacks (five sheets per stack), as depicted in Fig. 2. Note that the paper sample exhibits a progressive increase in thickness along the beam path across its transverse profile. Data were acquired under two conditions: (1) varying exposure times from 1 to 50 s per mask step, and (2) varying the X-ray source focal spot size among 7, 20, and 50 $\mu$m at a fixed 50 s exposure. For all acquisitions, a set of 10 images was captured by sequentially stepping the random mask as required by the retrieval algorithm.

For the analysis, the mean dark-field signal was plotted against the number of paper stacks. To ensure statistical independence between sampled points and eliminate spatial data redundancy caused by overlapping evaluation windows, the dark-field signal was evaluated across a sparse $8 \times 16$ grid of regions of interest with a 15-pixel separation (matching the window dimension) between points. To assess the statistical significance of the dark-field signal, error bars were calculated using the 95 $\%$ confidence interval. Algorithm performance was subsequently quantified by applying a linear regression to all data points. From the linear fit, sensitivity was defined as the slope, where a steeper value indicates a greater response to the small-angle X-ray scattering. Linearity was assessed by the coefficient of determination ($R^2$), with values approaching 1.0 signifying a strong linear response. Finally, signal bias was determined from the y-intercept, for which a value near zero is considered ideal as it represents a minimal background signal offset.

Finally, sample-induced spectral changes were assessed by acquiring X-ray energy spectra with a photon-counting detector while a paper phantom (0 to 5 stacks) was positioned in the beam path. To evaluate the linearity of the attenuation signal, a linear regression was established using only the baseline data points (0 to 2 stacks), where spectral filtration remains minimal. Residual analysis was then performed by calculating the deviation of each measured attenuation signal from the value predicted by this baseline linear fit across all thicknesses. To assess the statistical significance of these deviations, error bars were calculated using the 95 $\%$ confidence interval.

\begin{figure}[h!] 
    \centering 
    \begin{subfigure}{0.8\linewidth}
        \centering
        \includegraphics[width=\linewidth]{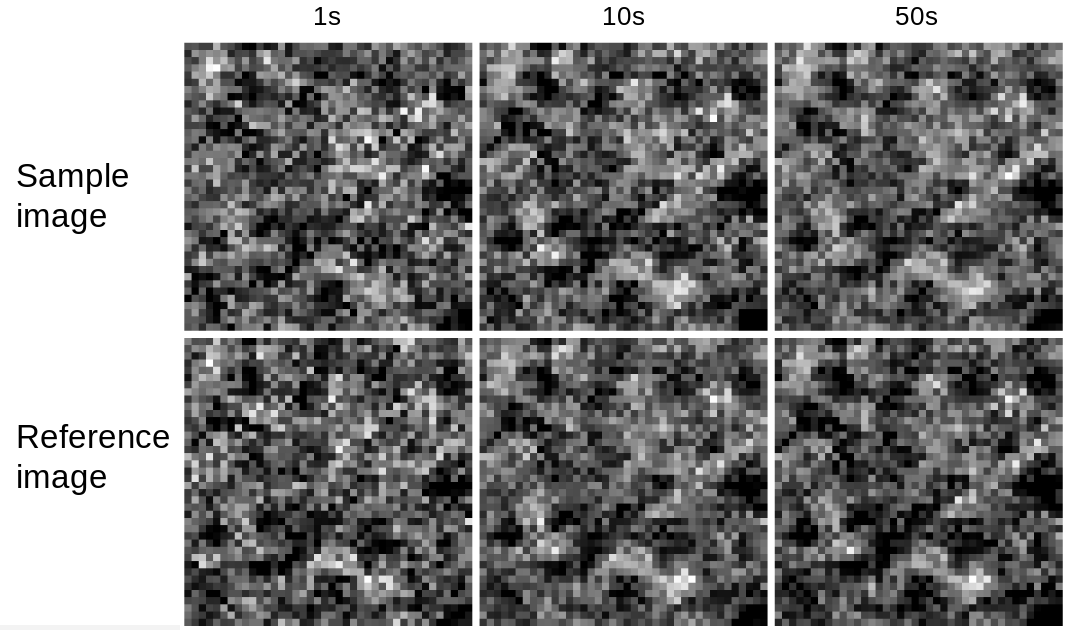} 
        \caption{} %
        \label{fig:3timeppt} 
    \end{subfigure}
    
    \begin{subfigure}{0.3\linewidth}
        \centering
        \includegraphics[width=\linewidth]{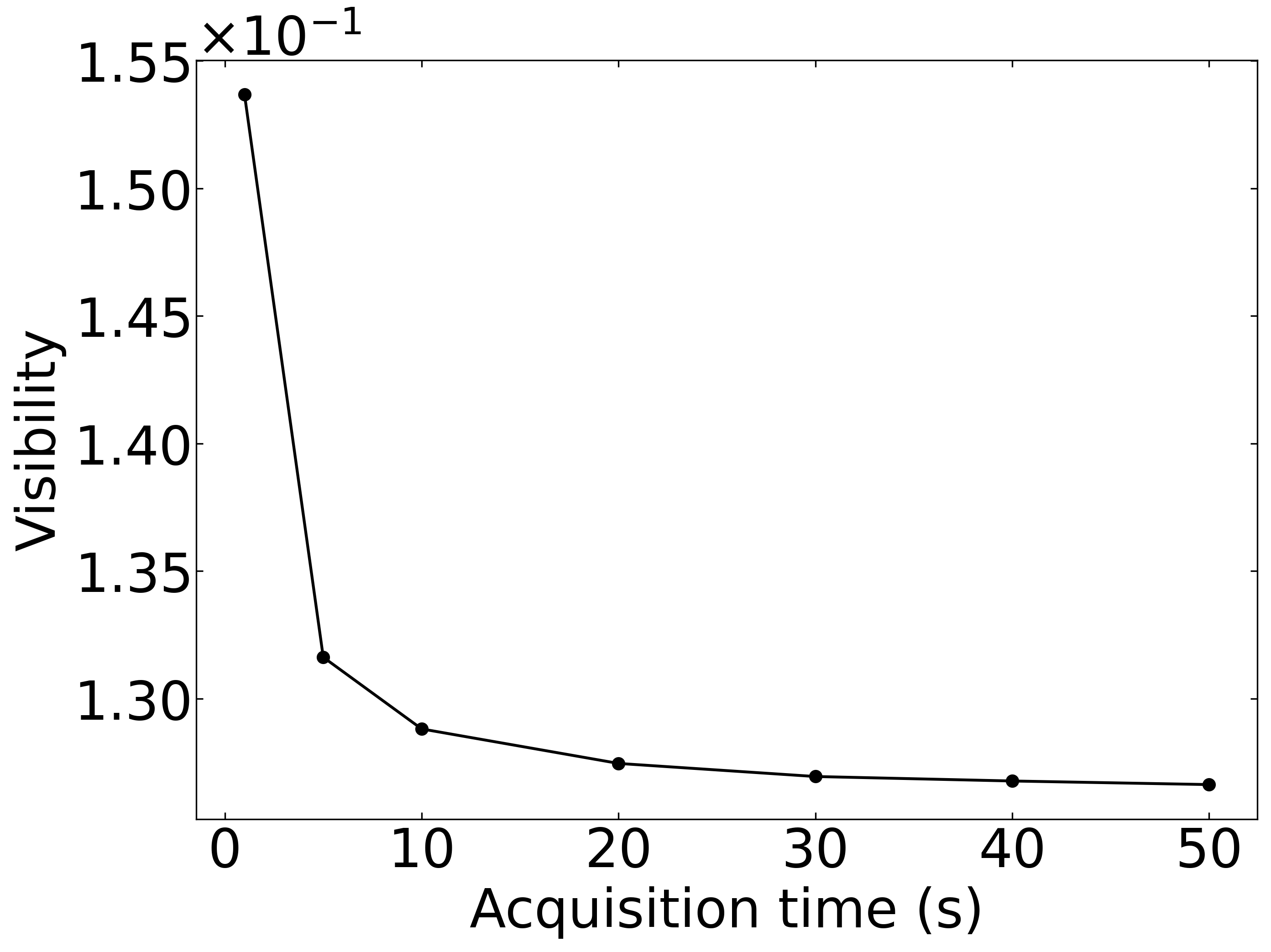} 
        \caption{} 
        \label{fig:3timev}
    \end{subfigure}
    \hfill
    \begin{subfigure}{0.3\linewidth}
        \centering
        \includegraphics[width=\linewidth]{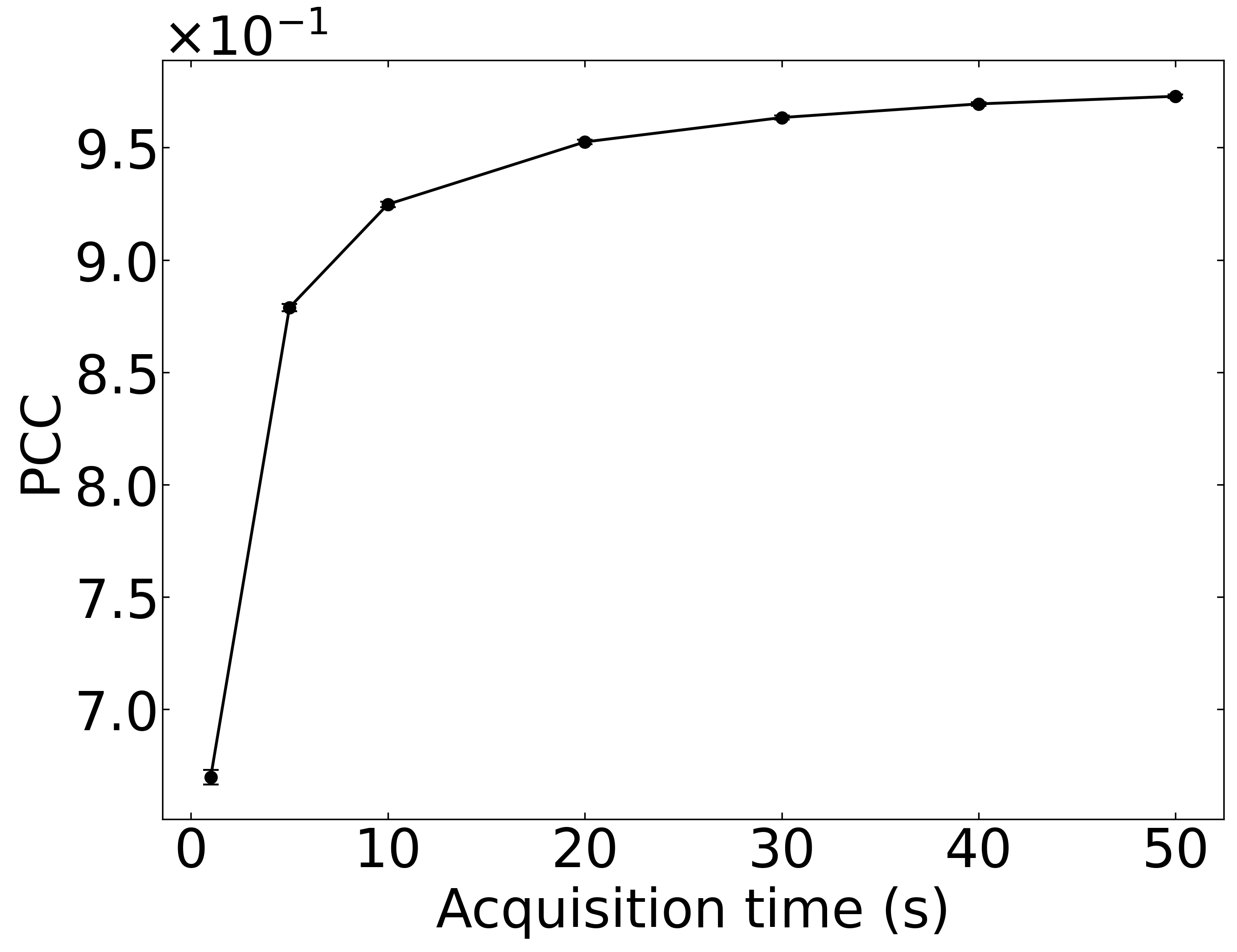} 
        \caption{} %
        \label{fig:3timepcc}  
    \end{subfigure}
    \hfill
    \begin{subfigure}{0.3\linewidth}
        \centering
        \includegraphics[width=\linewidth]{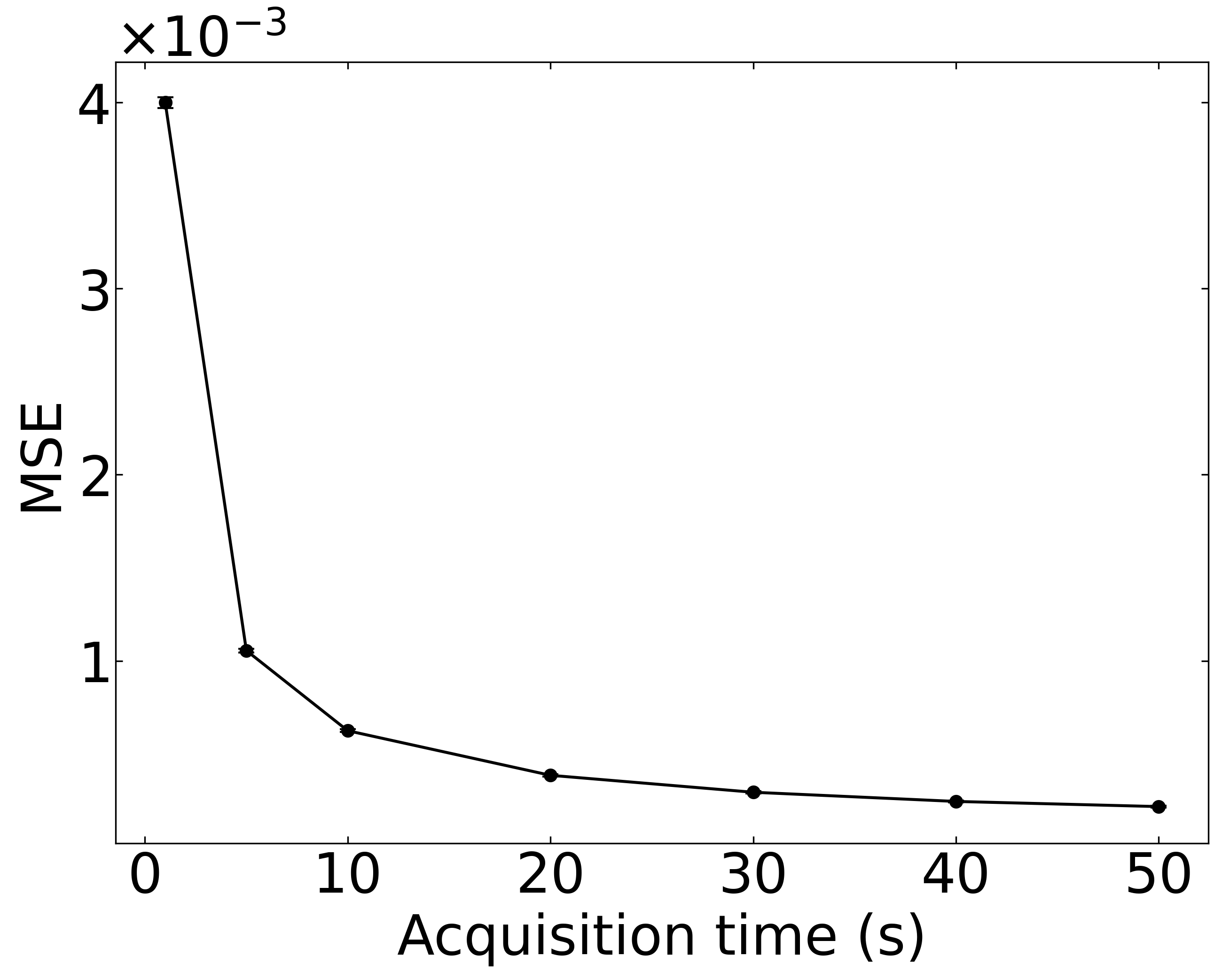} 
        \caption{}
        \label{fig:3timsmse} 
    \end{subfigure}       
    \caption{Effect of exposure time on image quality and quantitative metrics. (a) Representative sample (top) and reference (bottom) speckle images. (b) Visibility, (c) Pearson Correlation Coefficient (PCC), and (d) Mean Squared Error (MSE) are plotted as a function of exposure time, which ranges from 1 s to 50 s. Error bars indicate the 95$\%$ confidence interval.}
    \label{fig3} 
\end{figure}

\section{Results}  

\begin{figure}[h!] 
    \centering 
    \begin{subfigure}{0.8\linewidth}
        \centering
        \includegraphics[width=\linewidth]{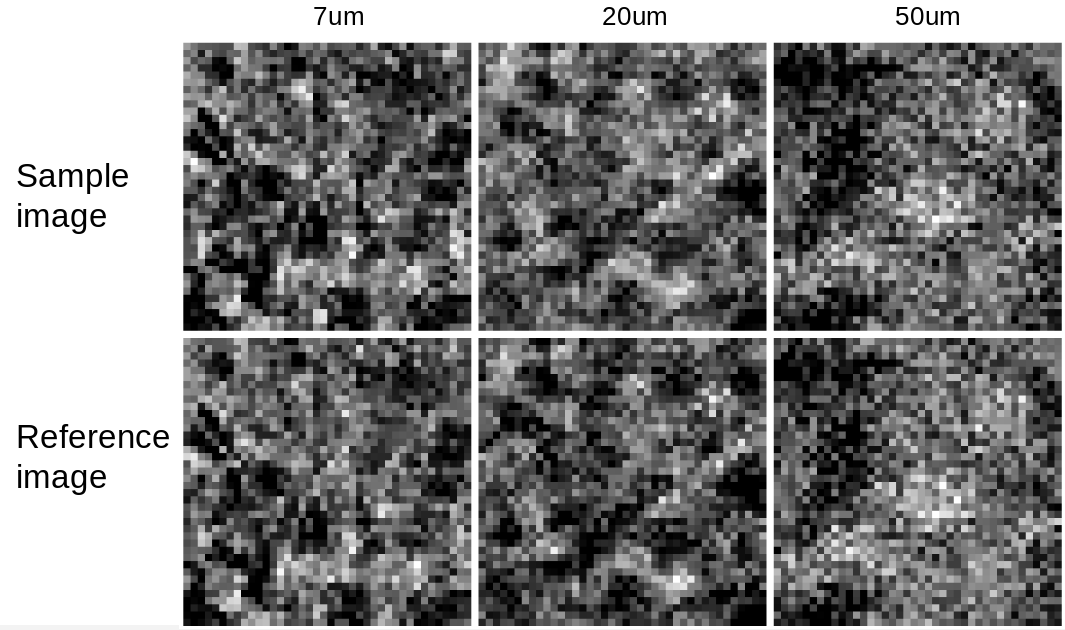} 
        \caption{} %
        \label{fig:4srcppt} 
    \end{subfigure}
    
    \begin{subfigure}{0.3\linewidth}
        \centering
        \includegraphics[width=\linewidth]{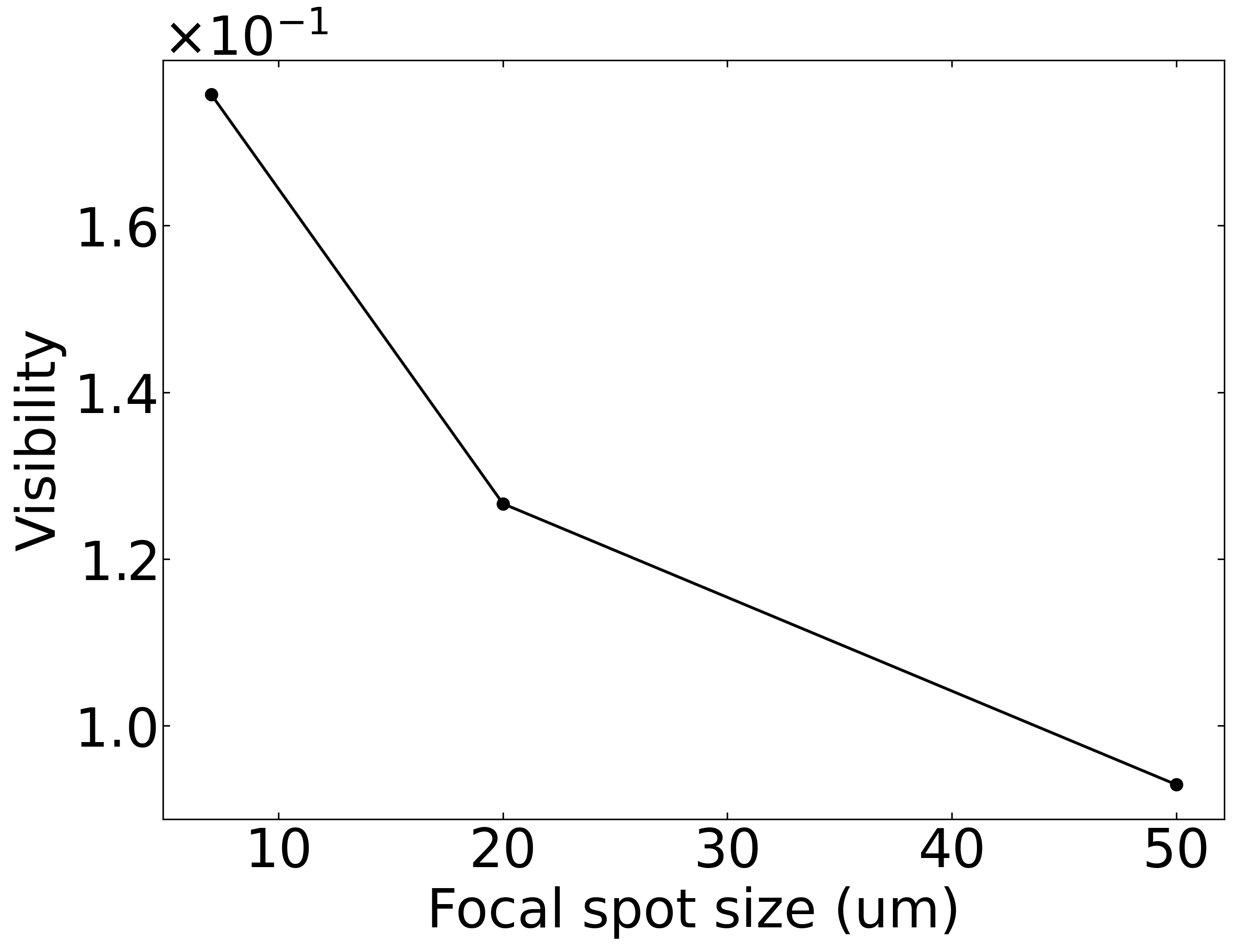} 
        \caption{} 
        \label{fig:4srcv}
    \end{subfigure}
    \hfill
    \begin{subfigure}{0.3\linewidth}
        \centering
        \includegraphics[width=\linewidth]{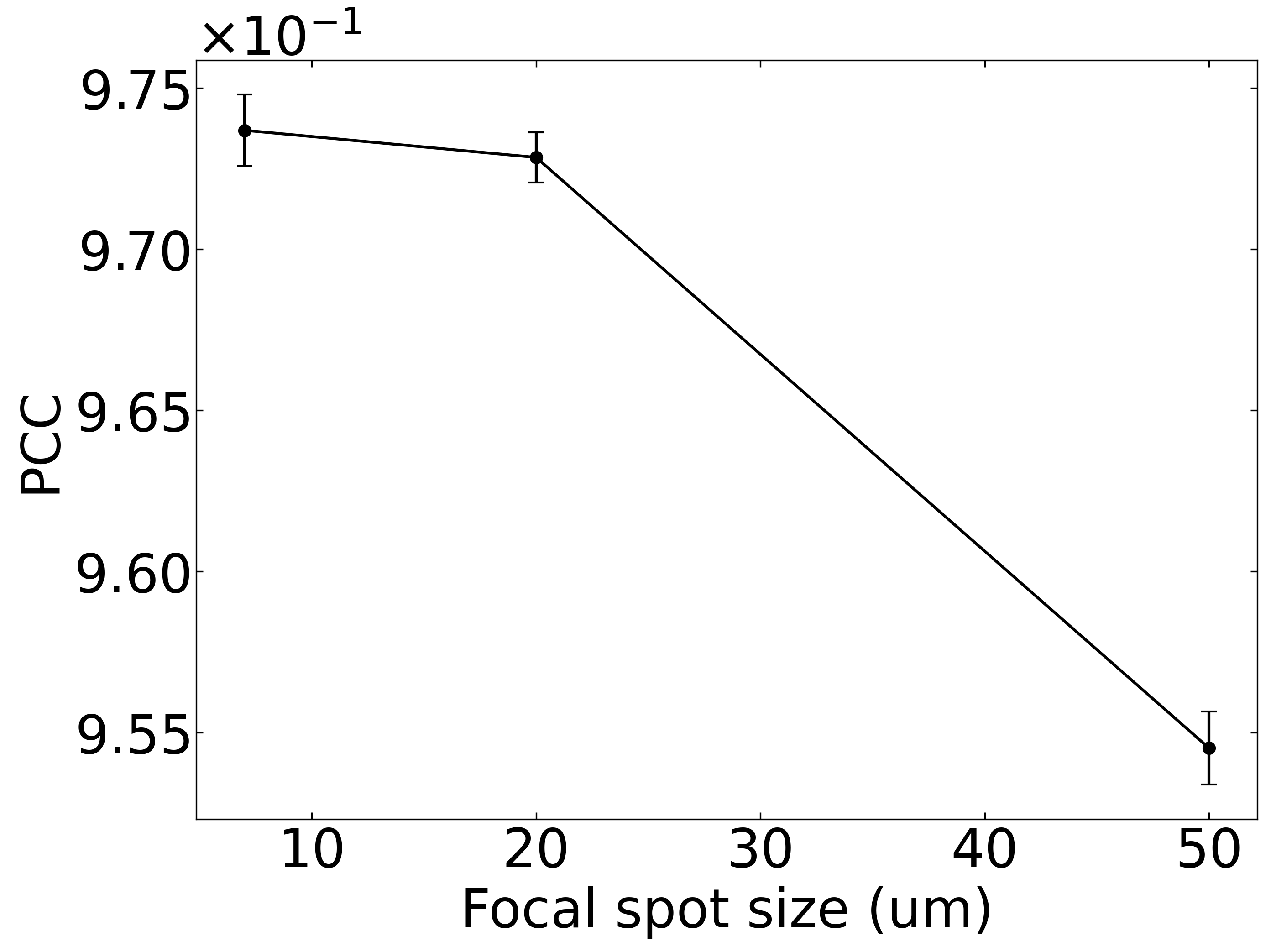} 
        \caption{} %
        \label{fig:4srcpcc}  
    \end{subfigure}
    \hfill
    \begin{subfigure}{0.3\linewidth}
        \centering
        \includegraphics[width=\linewidth]{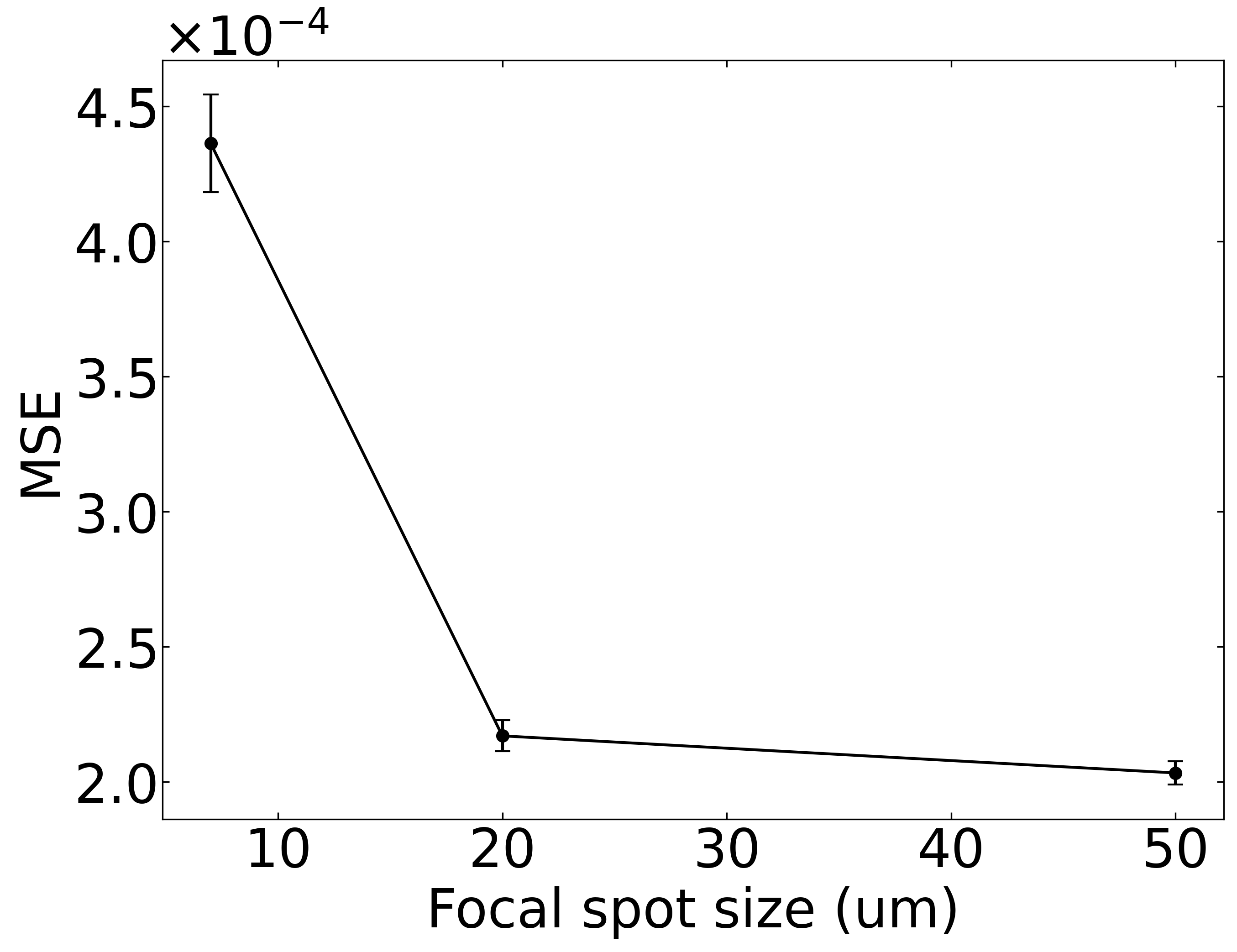} 
        \caption{}
        \label{fig:4srcsmse} 
    \end{subfigure}       
    \caption{Effect of X-ray source focal spot size on image quality and quantitative metrics. (a) Representative sample (top) and reference (bottom) speckle images. (b) Visibility, (c) Pearson Correlation Coefficient (PCC) and (d) Mean Squared Error (MSE) for focal spot sizes of 7, 20, and 50 $\mu$m. Error bars indicate the 95$\%$ confidence interval.}
    \label{fig4} 
\end{figure}

In this section, we analyze the effect of photon noise and source blurring on the speckle pattern and evaluate the performance of two representative dark-field retrieval algorithms.
\subsection{Effect on speckle pattern}
\subsubsection{Effect of photon noise on speckle pattern} 

To isolate the impact of photon statistics from source blur, the effect of photon noise was evaluated by varying the exposure time per mask step (directly altering the total collected photon counts per frame) while maintaining a fixed focal spot size of $20\ \mu\text{m}$. Figure 3 illustrates the representative speckle patterns (a) along with the corresponding quantitative metrics as a function of exposure time: speckle visibility (b), Pearson correlation coefficient (c), and mean squared error (d).
At short exposure times, the reference images exhibit an artificially elevated speckle visibility (Fig.3(b)). Specifically, as exposure time decreases below 10 $\text{s}$, the measured visibility increases sharply, reaching its peak at 1 $\text{s}$ before asymptotically approaching a stable value of approximately 0.13 as the exposure time extends toward 10 $\text{s}$ and beyond. This inflated visibility is an artifact caused by dominant photon shot noise, which artificially increases the local intensity standard deviation. This noise-driven instability is further corroborated by the similarity metrics in Fig. 3(c) and 3(d), where shorter exposure times yield a marked drop in the Pearson correlation coefficient and a corresponding spike in the mean squared error between the reference and sample images.

\subsubsection{Effect of source blurring on speckle pattern}
To isolate the influence of source blur from photon noise statistics, the effect of X-ray source blurring was evaluated while maintaining a constant total photon count (fixed exposure time of 50 $\text{s}$) across all steps. As demonstrated in Fig. 4(a) and (b), speckle visibility monotonically diminishes as the focal spot size expands from $7$ to $50\ \mu\text{m}$, directly reflecting the loss of spatial coherence and source blurring. 
In contrast, the image similarity metrics, Pearson correlation coefficient and mean squared error, exhibit relatively minor overall variations across the three focal spot sizes. However, a subtle trade-off is observed: the sharpest focal spot ($7\ \mu\text{m}$) yields a slightly higher mean squared error compared to larger focal spot sizes, despite maintaining a high Pearson correlation coefficient comparable to the $20\ \mu\text{m}$ condition. 

\subsection{Comparison of dark-field retrieval algorithms}
A direct performance comparison between the XST-XSVT and LCS algorithms is presented in Fig. 5 through 9, with quantitative metrics summarized in Table 1.

\subsubsection{Effect of photon noise on dark-field signal}
Figure 5 illustrates the retrieved attenuation and dark-field signals as a function of paper stack thickness—stacked incrementally along the X-ray beam propagation path, as illustrated in Fig.1—across varying exposure times. As exposure time decreased, a clear reduction in dark-field contrast was observed for both frameworks, indicating tracking performance degradation under elevated noise levels. Qualitative inspection of Fig. 5(a) and (b) reveals that at an exposure time of $1\,\text{s}$, the dark-field signal was effectively lost for both retrieval frameworks. 
These trends are quantitatively analyzed in Fig. 6(a) and (b) and summarized in Table 1(a). Here, signal sensitivity is defined as the slope of the linear fit, signal linearity is represented by the coefficient of determination ($R^2$), and signal bias is quantified by the $y$-intercept. As exposure time was reduced from $50\,\text{s}$ to $1\,\text{s}$, the sensitivity diminished for both methods, decreasing by $64.2\%$ (from $0.053$ to $0.019$) for XST-XSVT, and dropping precipitously by $91.3\%$ (from $0.023$ to $0.002$) for the LCS framework. Similarly, the linearity deteriorated under photon-limited conditions, falling by $37.4\%$ (from $0.99$ to $0.62$) for XST-XSVT and collapsing by $85.7\%$ (from $0.98$ to $0.14$) for LCS. Notably, the explicit XST-XSVT approach consistently maintained higher $R^2$ values, demonstrating superior robustness to noise compared to the LCS method. Another key divergence appeared in the signal bias: while the LCS intercept elevated from $0.06$ to $0.19$ as exposure time decreased—reflecting noise-induced spurious dark-field signals—the XST-XSVT intercept remained remarkably stable with negligible variation ($0.0043$ to $-0.021$).

\begin{figure}[ht!]   
    \centering 
    \begin{subfigure}{0.9\linewidth}
        \includegraphics[width=\linewidth]{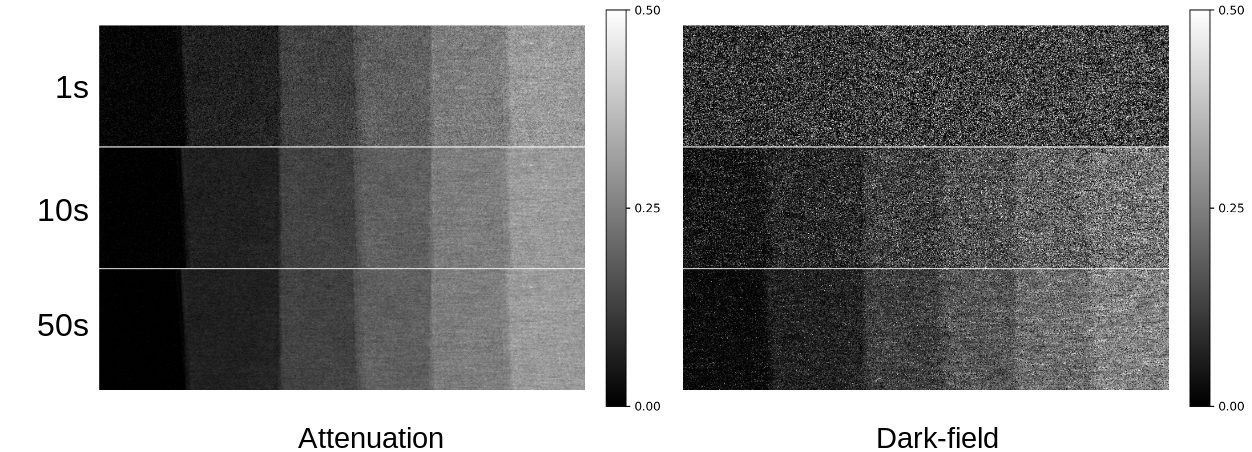} 
        \caption{XST-XSVT}
        \label{fig:5df} 
    \end{subfigure}
    \begin{subfigure}{0.9\linewidth} 
        \includegraphics[width=\linewidth]{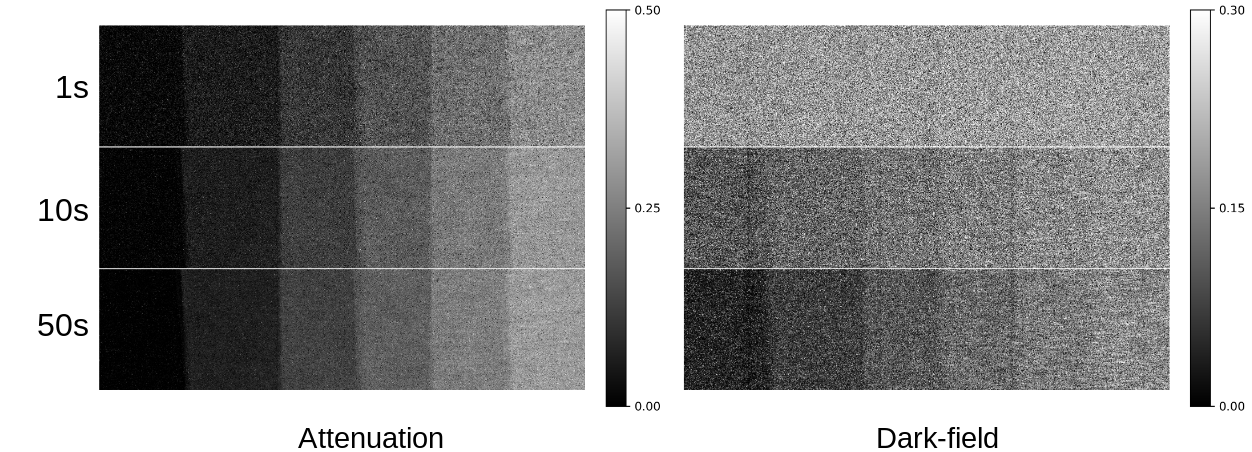} 
        \caption{LCS}
        \label{fig:5att} 
    \end{subfigure}
    \caption{Comparison of attenuation (left) and dark-field (right) signals, as a function of paper stack thickness, retrieved by the (a) XST-XSVT and (b) LCS  methods for various exposure times.}
    \label{fig5}
\end{figure}

\begin{figure}[ht!]   
    \centering 
    \begin{subfigure}{0.48\linewidth} 
        \includegraphics[width=\linewidth]{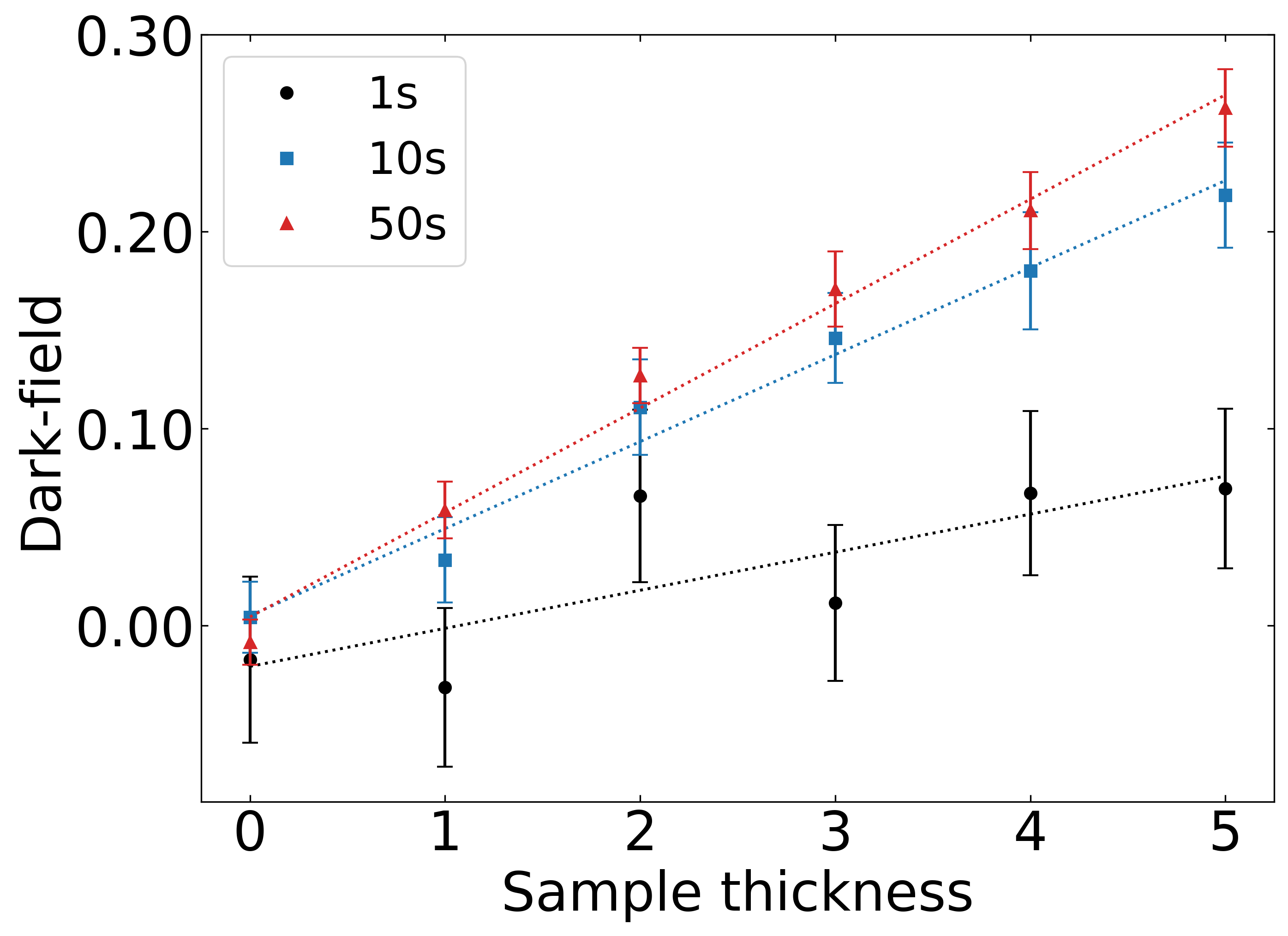} 
        \caption{}
        \label{fig:5linxsvttime} 
    \end{subfigure}
    \hfill
    \begin{subfigure}{0.48\linewidth} 
        \includegraphics[width=\linewidth]{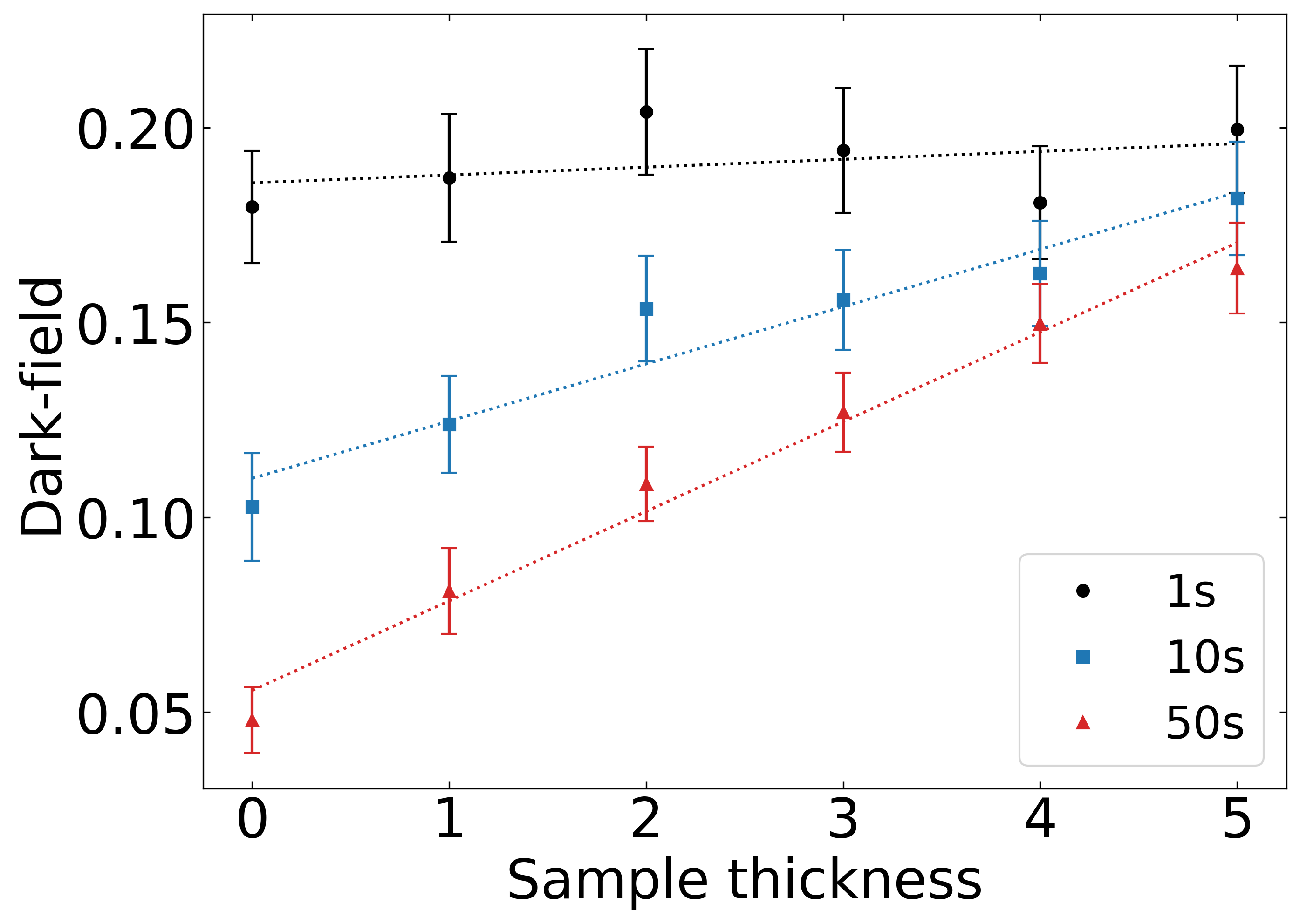} 
        \caption{}
        \label{fig:5linlcstime} 
    \end{subfigure}    
    \caption{Comparison of plots of the dark-field signal versus paper stack thickness for the (a) XST-XSVT and (b) LCS methods. Error bars indicate the 95$\%$ confidence interval.}
    \label{fig6}
\end{figure}

\begin{figure}[ht!] %
    \centering %
    \begin{subfigure}{0.9\linewidth} %
        \includegraphics[width=\linewidth]{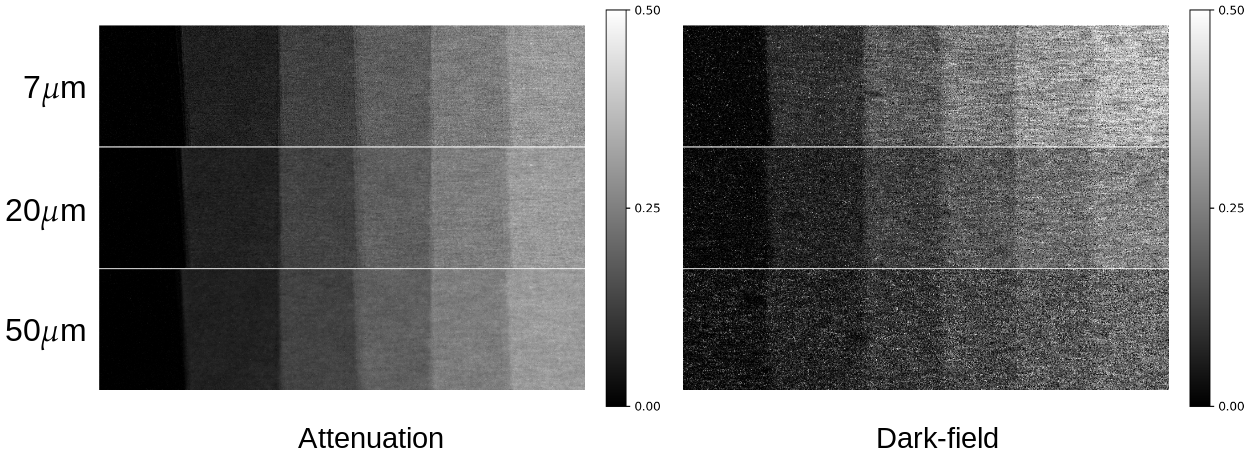}
        \caption{XST-XSVT}
        \label{fig:8dfsrcxsvt} %
    \end{subfigure} 
    \begin{subfigure}{0.9\linewidth} 
        \includegraphics[width=\linewidth]{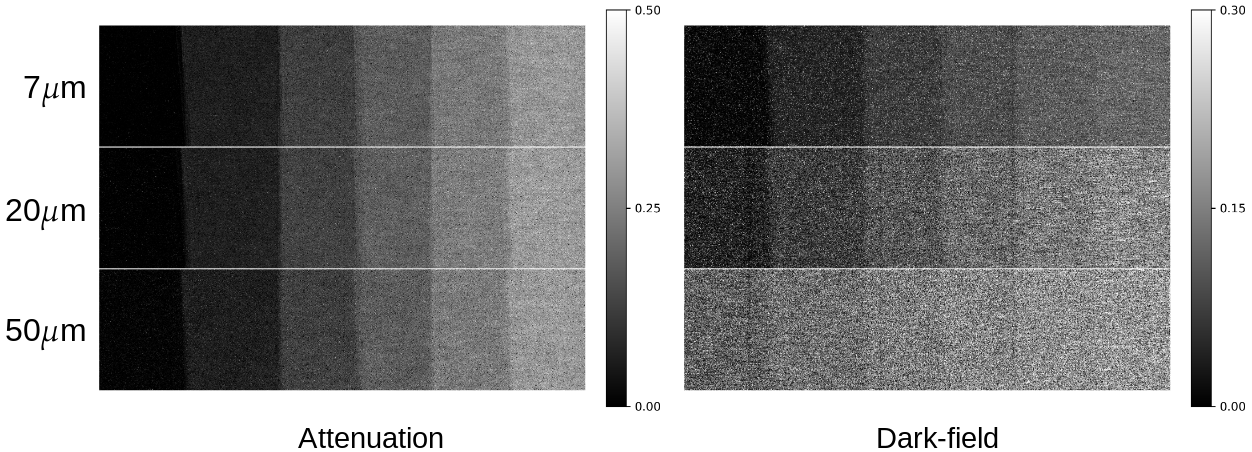} 
        \caption{LCS}
        \label{fig:8dfsrclcs}
    \end{subfigure} 
    \caption{Comparison of attenuation (left) and dark-field (right) images as a function of paper stack thickness, retrieved by the (a) XST-XSVT and (b) LCS methods for different focal spot sizes.}
    \label{fig8} 
\end{figure}

\begin{figure}[ht!] %
    \centering %
    \begin{subfigure}{0.48\linewidth} 
        \includegraphics[width=\linewidth]{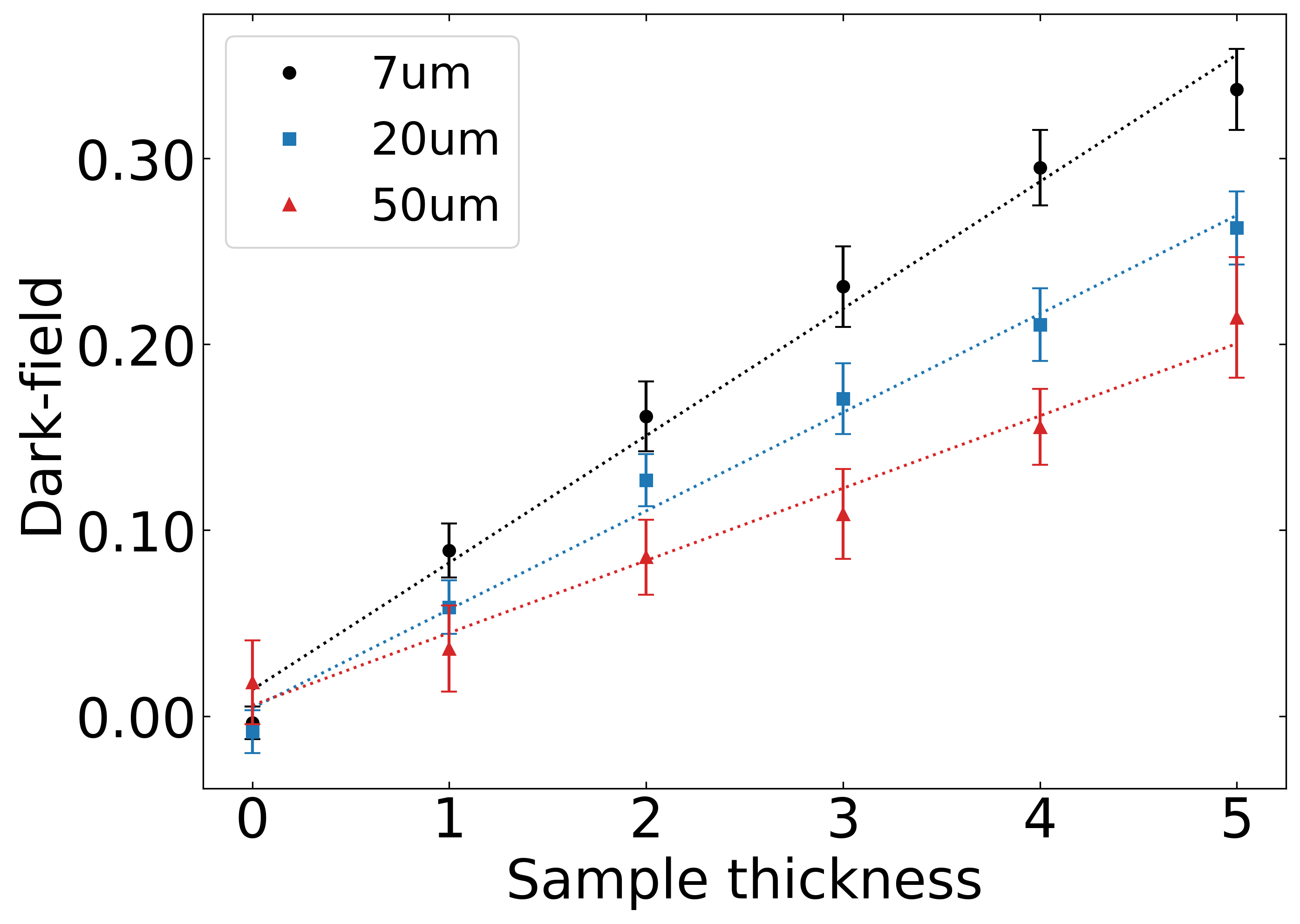} 
        \caption{}
        \label{fig:9attlinxsvtsrc}
    \end{subfigure}
    \hfill
    \begin{subfigure}{0.48\linewidth}
        \includegraphics[width=\linewidth]{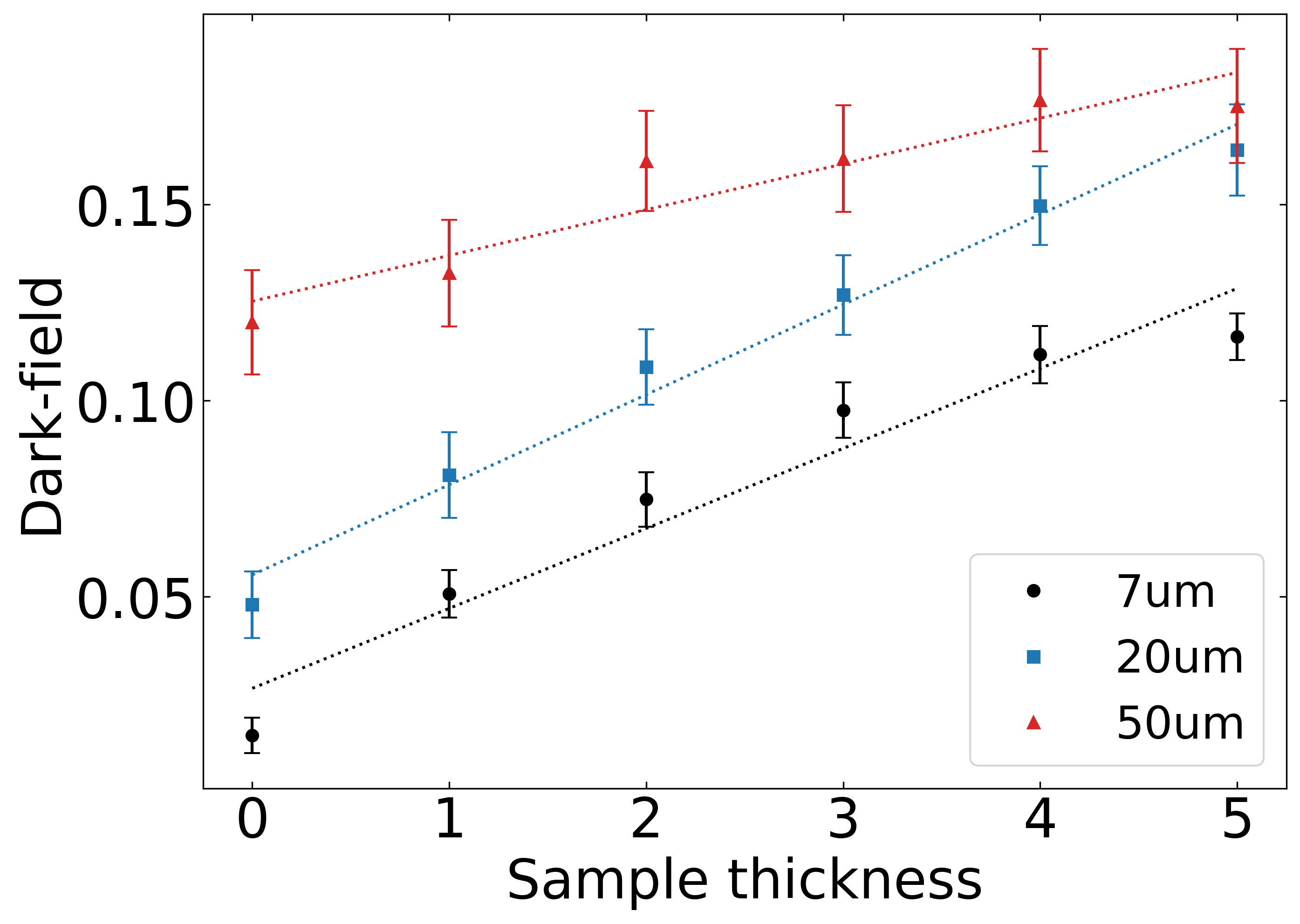} 
        \caption{}
        \label{fig:9attlinlcssrc}
    \end{subfigure}    
    \caption{Comparison of plots of the dark-field signal versus paper stack thickness for the (a) XST-XSVT and (b) LCS methods. Error bars indicate the 95$\%$ confidence interval.}
    \label{fig9} 
\end{figure}

\begin{table}[h!]
    \centering
    \caption{Linear regression analysis of dark-field signals: (a) exposure time and (b) focal spot size. Linearity, sensitivity, and bias are quantified by $R^2$, the slope, and the $y$-intercept of the linear regression fits, respectively.}
    \label{tab:combined_results}
    \vspace{0.5em} 

    \begin{subtable}[b]{0.48\linewidth}
        \centering
        \scalebox{0.72}{
        \begin{tabular}{|l|l|c|c|c|}
        \hline
        \textbf{Method} & \textbf{Time} & \textbf{R$^2$} & \textbf{Slope} & \textbf{Intercept} \\
        \hline
        \multirow{3}{*}{XST-XSVT} & 1s  & 0.62 & 0.019  & -0.021  \\
                                  & 10s & 0.98 & 0.044  & 0.0050   \\
                                  & 50s & 0.99 & 0.053  & 0.0043   \\
        \hline
        \multirow{3}{*}{LCS}      & 1s  & 0.14 & 0.002  & 0.19   \\
                                  & 10s & 0.93 & 0.015  & 0.11   \\
                                  & 50s & 0.98 & 0.023  & 0.06   \\
        \hline
        \end{tabular}
        }
    \end{subtable}
    \hfill 
    \begin{subtable}[b]{0.48\linewidth}
        \centering
        \scalebox{0.72}{
        \begin{tabular}{|l|l|c|c|c|}
        \hline
        \textbf{Method} & \textbf{Focal Spot} & \textbf{R$^2$} & \textbf{Slope} & \textbf{Intercept}  \\
        \hline
        \multirow{3}{*}{XST-XSVT} & 7$\mu$m  & 0.99 & 0.068 & 0.014    \\
                                  & 20$\mu$m & 0.99 & 0.053 & 0.004  \\
                                  & 50$\mu$m & 0.98 & 0.039 & 0.006    \\
        \hline
        \multirow{3}{*}{LCS}      & 7$\mu$m  & 0.94 & 0.020 & 0.03   \\
                                  & 20$\mu$m & 0.98 & 0.023 & 0.06   \\
                                  & 50$\mu$m & 0.89 & 0.012 & 0.13   \\
        \hline
        \end{tabular}
        }
    \end{subtable}
\end{table}

\begin{figure}[h!]
    \centering
    \begin{subfigure}{0.5\linewidth} 
        \centering
        \includegraphics[width=\linewidth]{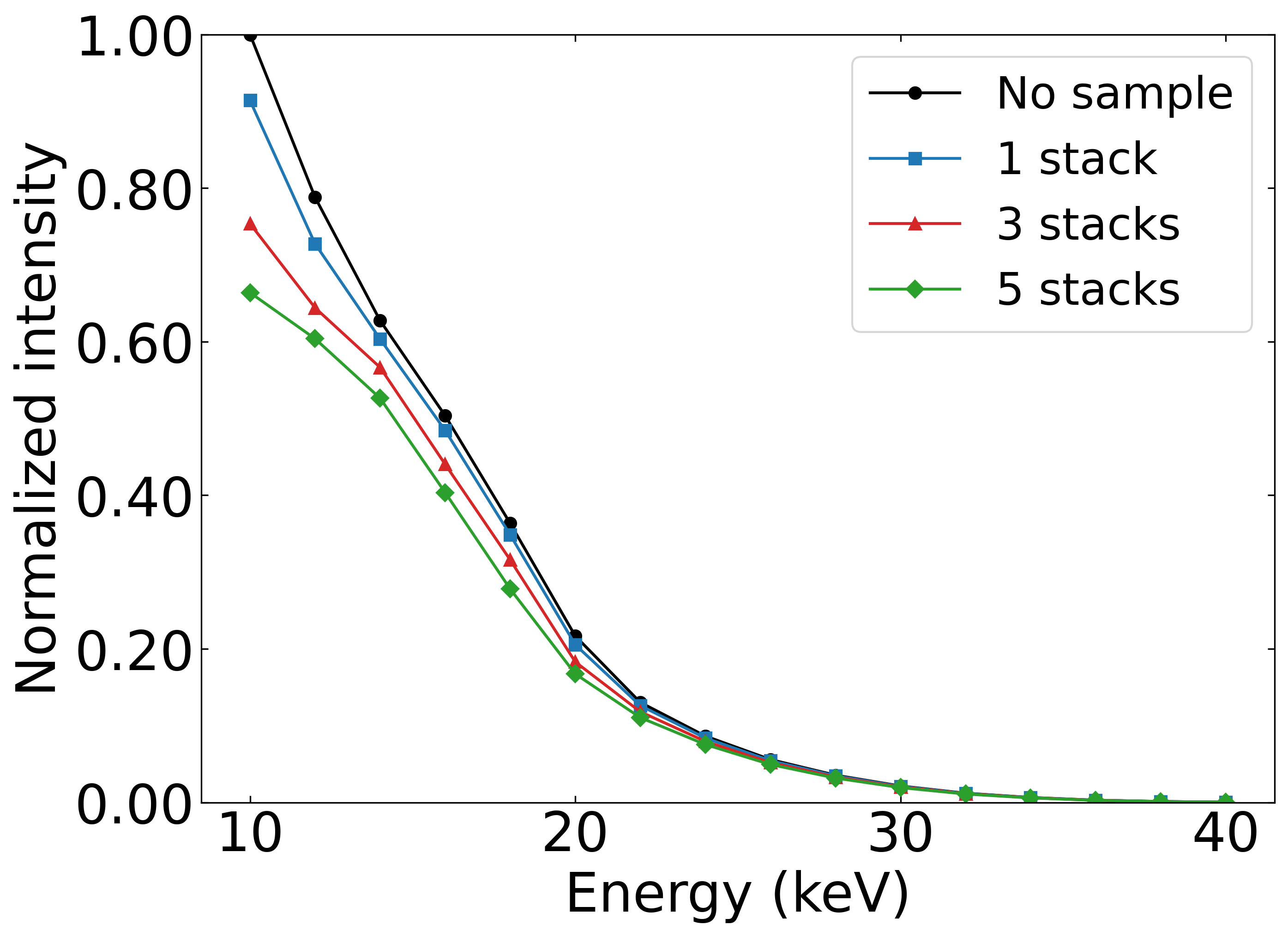} 
        \caption{}
        \label{fig:spectrum}
    \end{subfigure} 

    \vspace{2em} 

    \begin{subfigure}{0.48\linewidth}
        \centering
        \includegraphics[width=\linewidth]{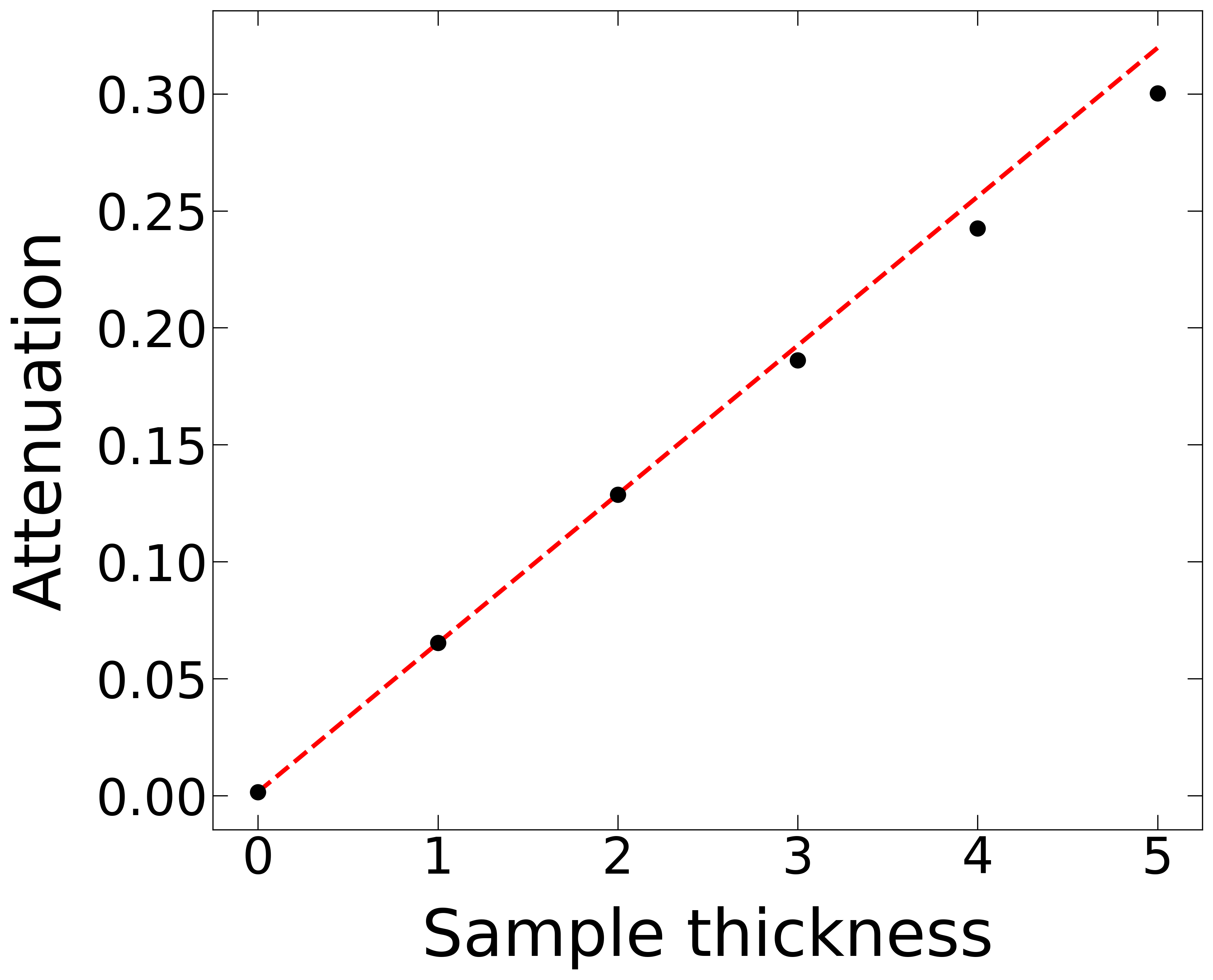} 
        \caption{}
        \label{fig:att}
    \end{subfigure}
    \hfill
    \begin{subfigure}{0.48\linewidth}
        \centering
        \includegraphics[width=\linewidth]{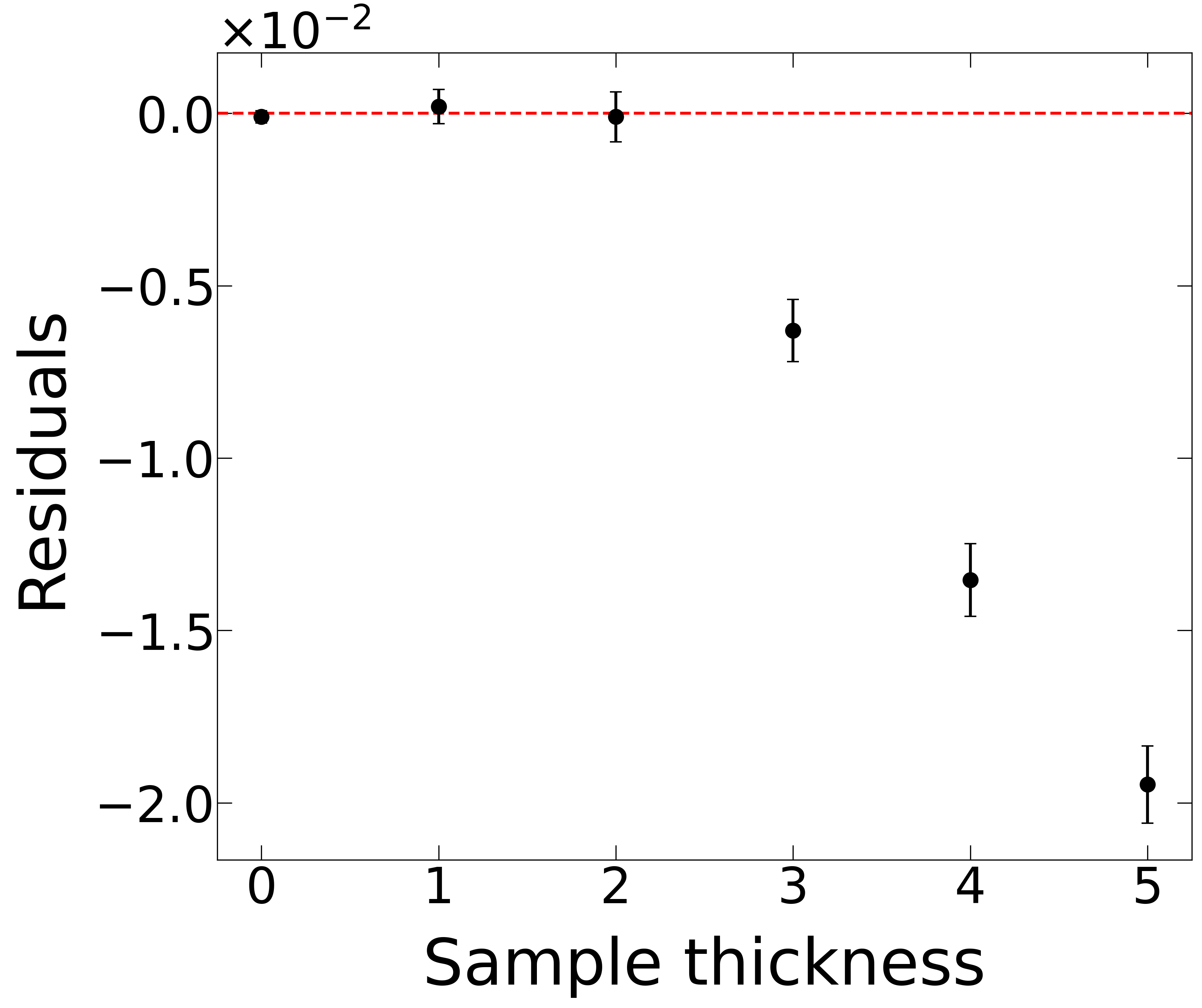} 
        \caption{}
        \label{fig:residuals}
    \end{subfigure}    

    \caption{Spectral and attenuation analysis: (a) raw X-ray spectra; (b) attenuation signal versus paper thickness; and (c) residual analysis. Error bars indicate the 95$\%$ confidence interval.}
    \label{fig10}     
\end{figure}

\subsubsection{Effect of source blurring on dark-field signal}
Figure 7 presents the retrieved dark-field signal as a function of total paper stack thickness across varying focal spot sizes. Similar to the trends observed with decreasing exposure times, increasing the focal spot size led to a systematic reduction in dark-field contrast, signaling a degradation in tracking performance due to source blurring. 
Qualitative inspection of Fig. 7 shows that at a $50\,\mu\text{m}$ focal spot, the dark-field signal becomes nearly imperceptible for both retrieval frameworks, marking the upper limit of source-induced blurring for reliable signal extraction in this setup.

These trends are quantitatively analyzed in Fig. 8(a) and (b) and summarized in Table 1(b). The two algorithms exhibited divergent responses to variations in focal spot size. For the explicit XST-XSVT approach, dark-field sensitivity was inversely related to focal spot size, decreasing by $42.6\%$ (from $0.068$ to $0.039$) as the focal spot size expanded from $7\,\mu\text{m}$ to $50\,\mu\text{m}$; as expected, peak sensitivity was attained with the smallest ($7\,\mu\text{m}$) focal spot. Furthermore, its linearity remained robust, exhibiting a negligible $1.0\%$ drop in $R^2$ (from $0.99$ to $0.98$). In contrast, the intrinsic LCS method displayed a non-monotonic response, where the $20\,\mu\text{m}$ focal spot yielded both higher sensitivity ($0.023$) and superior linearity ($0.98$) compared to the $7\,\mu\text{m}$ source. As the focal spot size increased further to $50\,\mu\text{m}$, LCS's linearity degraded by $5.3\%$ (falling from $0.94$ to $0.89$) and its sensitivity dropped by $47.8\%$ relative to its peak value at $20\,\mu\text{m}$ (from $0.023$ to $0.012$). A significant divergence was also noted in the signal bias: the LCS intercept elevated from $0.03$ to $0.13$ with increasing focal spot size, whereas XST-XSVT recorded its maximum of $0.014$ at $7\,\mu\text{m}$ and a minimum of $0.004$ at $20\,\mu\text{m}$. Consistent with the noise analysis, the substantial background bias of the LCS method persisted across all focal spot sizes.

As observed in the quantitative results presented in Fig. 6 and 8, a sublinear behavior becomes apparent as the sample thickness increases. This trend is attributed to a beam-hardening effect, the physical origin of which is further elucidated through spectral analysis. As shown in Fig. 9(a), measurements verify a distinct hardening of the beam, where the mean photon energy shifts from $17.69\,\text{keV}$ to $18.88\,\text{keV}$ as the filtration reaches 5 stacks. This spectral shift provides the physical context for the signal impairment observed in our algorithms. While the attenuation signal follows the linear fit established in Fig. 9(b) for thinner samples, the beam hardening causes the measured values to drop below the predicted baseline for thicker objects. This is most clearly visualized in the residual plot in Fig. 9(c), where a non-zero divergence emerges specifically beyond the 3-stack threshold. Note that the 95 $\%$ confidence interval for the attenuation signal are smaller than the marker size and are therefore not shown.


\section{Discussion}
This study demonstrates that photon noise and source blurring significantly degrade the dark-field linearity and sensitivity while introducing spurious signal bias. This degradation is driven by a structural mismatch between reference and sample speckle patterns, quantified here by increased mean squared error and decreased Pearson correlation coefficient. 

Crucially, the statistical metrics employed—visibility, the Pearson correlation coefficient, and the mean squared error distinct sensitivities depending on whether the system is photon-limited or blur-limited. While high speckle visibility is fundamentally required for accurate signal reconstruction \cite{zdora2018state}, our findings reveal its limited capacity to capture localized speckle quality under low-photon conditions; notably, visibility failed to track the progressive dark-field degradation induced by increasing noise. In contrast, the Pearson correlation coefficient and the mean squared error more reflected this noise-induced degradation, with the Pearson correlation coefficient decreasing and the mean squared error increasing systematically as exposure times were reduced.


When assessing the impact of source blurring under sufficient photon flux (where total photon counts per frame were kept constant across all conditions), a contrary trend emerges: speckle visibility decreased with increasing focal spot size as theoretically expected, whereas variations in the Pearson correlation coefficient and the mean squared error were far less pronounced than those observed in the photon-limited regime. Intriguingly, while the smallest focal spot ($7\ \mu\text{m}$) yielded the highest speckle visibility, it simultaneously produced the highest mean squared error and a Pearson correlation coefficient comparable to the $20\ \mu\text{m}$ source.
This anomaly is attributed to an interplay between high-frequency spatial preservation, detector sampling limits, and microscopic system vibrations. Because the $7\ \mu\text{m}$ focal spot minimizes source blurring (blurring of $\sim 4.8\ \mu\text{m}$ at the sample plane), it preserves steep speckle gradients that are significantly narrower than the effective detector pixel pitch ($\sim 17\ \mu\text{m}$). This spatial undersampling renders the system acutely sensitive to sub-pixel mechanical vibrations \cite{pil2022direct}. As these sharp gradients oscillate across discrete pixel boundaries, they induce pixel jitter and generate localized intensity errors that artificially drive up the mean squared error. Conversely, the $20\ \mu\text{m}$ focal spot effectively functions as an optimal anti-aliasing filter; its source blurring ($\sim 13.8\ \mu\text{m}$) closely matches the effective pixel pitch, thereby dampening the impact of mechanical jitter, stabilizing the digitized pattern, and preserving the Pearson correlation coefficient. In contrast, the $50\ \mu\text{m}$ focal spot introduces excessive blurring ($\sim 34.6\ \mu\text{m}$) that physically erases high-frequency speckle structures, resulting in the lowest Pearson correlation coefficient. To substantiate this mechanism, a numerical simulation was conducted, the results of which closely align with our experimental data (provided in the Supplementary Information). These raw pattern distortions directly dictate the numerical stability and accuracy of subsequent dark-field signal retrieval.

The two physical constraints—photon noise and source blurring—affected both retrieval algorithms; however, the impact on the LCS method was notably more pronounced. The enhanced resilience of the XST-XSVT method is particularly relevant for clinical translation, where radiation dose constraints and limited acquisition windows are primary considerations. This performance gap stems from the inverse formulations of the two retrieval strategies. While LCS relies on an unregularized, point-wise algebraic inversion governed by the second-order Laplacian operator, XST-XSVT computes statistical variance ratios over a 2D spatial window.
Under short exposure times (Table 1(a)), applying the second-order Laplacian directly to photon-starved frames amplifies photon noise. Because LCS inverts these noisy derivatives independently at each pixel, the amplified random error shifts the baseline, generating a spurious background offset and reducing the signal linearity. Conversely, XST-XSVT evaluates intensity variance over a local ensemble of pixels. This spatial windowing inherently averages out noise, suppressing baseline offset and maintaining high response linearity even down to a $1\,\text{s}$ exposure.
These mathematical structures also dictate how each algorithm handles source-induced blurring (Table 1(b)). For XST-XSVT, expanding the focal spot causes a steady, monotonic decrease in signal sensitivity (slope) while preserving linearity ($R^2 \ge 0.98$) and negligible background bias. Its windowed variance calculation scales smoothly with reduced speckle contrast without introducing numerical distortions. In contrast, LCS displays a non-monotonic response driven by the raw pattern characteristics described above. Specifically, the jitter-induced undersampling errors at $7\,\mu\text{m}$ reduce its linearity and sensitivity, whereas the anti-aliasing stabilization at $20\,\mu\text{m}$ recovers peak linearity and sensitivity. Beyond this optimal point, severe blurring at $50\,\mu\text{m}$ suppresses the spatial Laplacian operator. Because this spatial gradient term acts as the direct scaling factor in the point-wise inversion, its dampening renders the inversion acutely ill-conditioned. Solving such an ill-conditioned system in the presence of residual noise induces numerical instability, ultimately suppressing signal sensitivity, compromising response linearity, and elevating background bias.

In this study, the LCS framework was evaluated in its native, unregularized form to assess the intrinsic physical limits and baseline numerical stability of its governing equations under photon noise and source blurring. Evaluating the unregularized framework prevents hyperparameter bias and isolates the intrinsic numerical stability of the point-wise formulation. Nevertheless, the limitations of LCS under noise and blur conditions highlight the clear necessity of stabilization, mirroring approaches such as Multimodal Intrinsic Speckle-Tracking \cite{alloo2023m}. These identified limitations provide a diagnostic basis for developing spatially adaptive regularization schemes in future studies.

Although both retrieval frameworks suffered from reduced linearity with increased sample thickness, the spectral shifts shown in Fig. 9(a) identify beam hardening as the underlying cause. That these non-linearities manifest identically in both dark-field and attenuation channels confirms that the deviation is a physical artifact of the source’s polychromatic nature. These findings underscore the necessity of beam hardening correction to ensure quantitative accuracy, consistent with observation in grating interferometry \cite{chabior2011beam}. Importantly, even in the presence of beam hardening, XST-XSVT consistently maintains higher linearity than LCS, further validating its advantage when utilizing laboratory X-ray sources.

Linear regression analysis was employed to evaluate the dark-field signal. Because artifacts like beam hardening distort the retrieved signal at greater thicknesses, evaluating the linear fit across all six sample thickness data points incorporates these spectral shifts. Notably, the overall fitting trends across the six data points align closely with those obtained from the first three points (which remain largely free from beam hardening). The primary discrepancy appears in the intercept for the $7\,\mu\text{m}$ focal spot size in XST-XSVT (Table 1(b)), where the relative elevation reflects the specific impact of beam hardening.

This study has certain limitations. The experiments were performed using a polychromatic, low-energy X-ray source, a thin phantom, and a high-resolution photon-counting detector. Therefore, directly extrapolating these findings to clinical settings—which involve high-energy spectra, thick biological tissues, and energy-integrating detectors with larger pixel sizes—requires caution. The interplay between beam hardening, Compton scatter, and varying pixel-to-speckle size ratios in clinical scenarios may alter the observed performance trade-offs.

These results define the physical and algorithmic constraints of laboratory-scale dark-field imaging, a necessary step toward clinical translation. The vulnerabilities identified in current tracking algorithms underscore the need for more robust signal retrieval frameworks. To overcome these limitations, future work will integrate our light transport models with the multi-energy binning capabilities of photon-counting detectors \cite{yuan2024transport,yuan2025single,vespucci2018robust,gursoy2013single,vazquez2020quantitative,das2014approximated}. Other benefits of using these advanced detectors would be in extracting spectrally dependent information with scatter correction \cite{lewis2019spectral,lewis2022energy} and possible understanding on anatomical and quantum noise \cite{kavuri2020relative}. This expansion is expected to circumvent the physical constraints of conventional speckle imaging, thereby broadening its diagnostic utility.

\section{Conclusion}
This study characterized the impact of photon noise and source blurring—primary physical constraints of laboratory X-ray tubes—on speckle-based dark-field retrieval. These physical limitations induce speckle pattern mismatches between reference and sample acquisitions, resulting in reduced retrieval sensitivity, loss of response linearity, and elevated background bias. Our findings demonstrate that the severity of this performance degradation is governed by the mathematical architecture of the retrieval algorithm. While the point-wise, differential-based LCS framework is susceptible to noise amplification and blur-induced ill-conditioning, the patch-wise XST-XSVT framework provides implicit spatial regularization through variance calculations, maintaining high response linearity and minimal signal bias. These insights establish a quantitative baseline for laboratory-scale speckle imaging. To address remaining physical artifacts such as beam hardening and advance clinical translation, future work will focus on integrating multi-energy photon-counting detection with advanced light transport modeling within the speckle formalism.

\section*{Funding.}
This work was partially supported by funding from from the NIH National Institute of Biomedical Imaging and Bioengineering (NIBIB) grant R01 EB EB029761, the US Department of Defense (DOD) Congressionally Directed Medical Research Program (CDMRP) Breakthrough Award BC151607 and the National Science Foundation CAREER Award 1652892.

\section*{Disclosures.}
The authors declare no conflicts of interest.

\section*{Data availability.}
Data underlying the results presented in this paper are not publicly available at this time but may be obtained from the authors upon reasonable request.

 
\bibliographystyle{unsrt} 
\bibliography{ref}

\clearpage
\appendix
\renewcommand{\thesection}{S\arabic{section}}
\renewcommand{\thefigure}{S\arabic{figure}}
\renewcommand{\thetable}{S\arabic{table}}
\renewcommand{\theequation}{S\arabic{equation}}
\setcounter{section}{0}
\setcounter{figure}{0}
\setcounter{table}{0}
\setcounter{equation}{0}

\clearpage
\begin{center}
    \textbf{\LARGE Supplementary Material}\\[0.5em]
\end{center}
\vspace{1em}

\section{Simulation Methodology}
To further elucidate the findings reported in the main text—specifically the observation that the smallest focal spot size yielded the highest MSE while maintaining a PCC comparable to the $20\ \mu\text{m}$ source—we conducted a numerical analysis using simulated jittered speckle patterns. This supplementary study aims to characterize the interaction between a stochastic scattering surface, focal spot size, and detector-level discretization under mechanical instability.

\subsection{Stochastic Surface Modeling}
To simulate a realistic scattering medium, we employ a multi-scale spatial correlation model. The process begins with the generation of a high-resolution 2D array of uncorrelated white noise, denoted as $I_{raw}(x, y)$.

To introduce physical correlation lengths representing different grain sizes, we apply a series of Gaussian filters to the raw noise. The multi-scale base pattern $P(x, y)$ is defined as the summation of three distinct spatial regimes:
\begin{equation}
    P(x, y) = \sum_{i \in \{2, 6, 15\}} G(x, y; \sigma_i) * I_{raw}(x, y)
\end{equation}
where $G(x, y; \sigma_i)$ represents a Gaussian kernel with standard deviation $\sigma_i$. These values correspond to fine microscopic grit ($\sigma=2$), nominal macroscopic grains ($\sigma=6$), and large-scale structural clustering ($\sigma=15$).

\subsection{Focal Spot Blurring}
The optical system is modeled as a low-pass filter defined by the focal spot size. The effective optical field $I_{opt}$ is obtained by convolving the base pattern with a Gaussian point spread function (PSF):
\begin{equation}
    I_{opt} = P(x, y) * G(x, y; \sigma_{spot})
\end{equation}
Following experimental approximations, the standard deviation of the blur kernel is related to the spot size by $\sigma_{spot} = \text{spot\_size} \times \frac{3}{7}$.

\subsection{Detector Discretization and Jitter}
The high-resolution optical field is discretized to match the detector pixel pitch through a spatial integration process. For an effective pixel size $D$, the detected intensity $I_{det}$ is calculated by averaging the high-resolution grid within a $D \times D$ window.

Mechanical jitter is introduced as a discrete spatial shift $\Delta x$ applied to the high-resolution field prior to the downsampling process. This allows for the simulation of sub-pixel vibrations, where $\Delta x < D$.

\section{Parameter Selection and Experimental Alignment}
The simulation parameters were specifically selected to maintain parity with the experimental configuration:

\begin{itemize}
    \item \textbf{Effective Pixel Size ($17\,\mu\text{m}$):} Set to match the detector pitch used in high-resolution data acquisition.
    \item \textbf{Focal Spot Sweep ($5\,\mu\text{m}$ to $60\,\mu\text{m}$):} Selected to characterize the transition from the diffraction-limited regime to the blur-dominated regime.
    \item \textbf{Mechanical Jitter ($2\,\mu\text{m}$):} Introduced to simulate typical sub-pixel mechanical instabilities observed in the experimental setup.
    \item \textbf{Additive Noise ($\sigma \in [0, 0.01]$):} Additive White Gaussian Noise (AWGN) was incorporated to evaluate the robustness of correlation metrics against varying signal-to-noise ratios (SNR).
\end{itemize}

\section{Results}
\begin{figure}[t]
    \centering
    \includegraphics[width=0.9\linewidth]{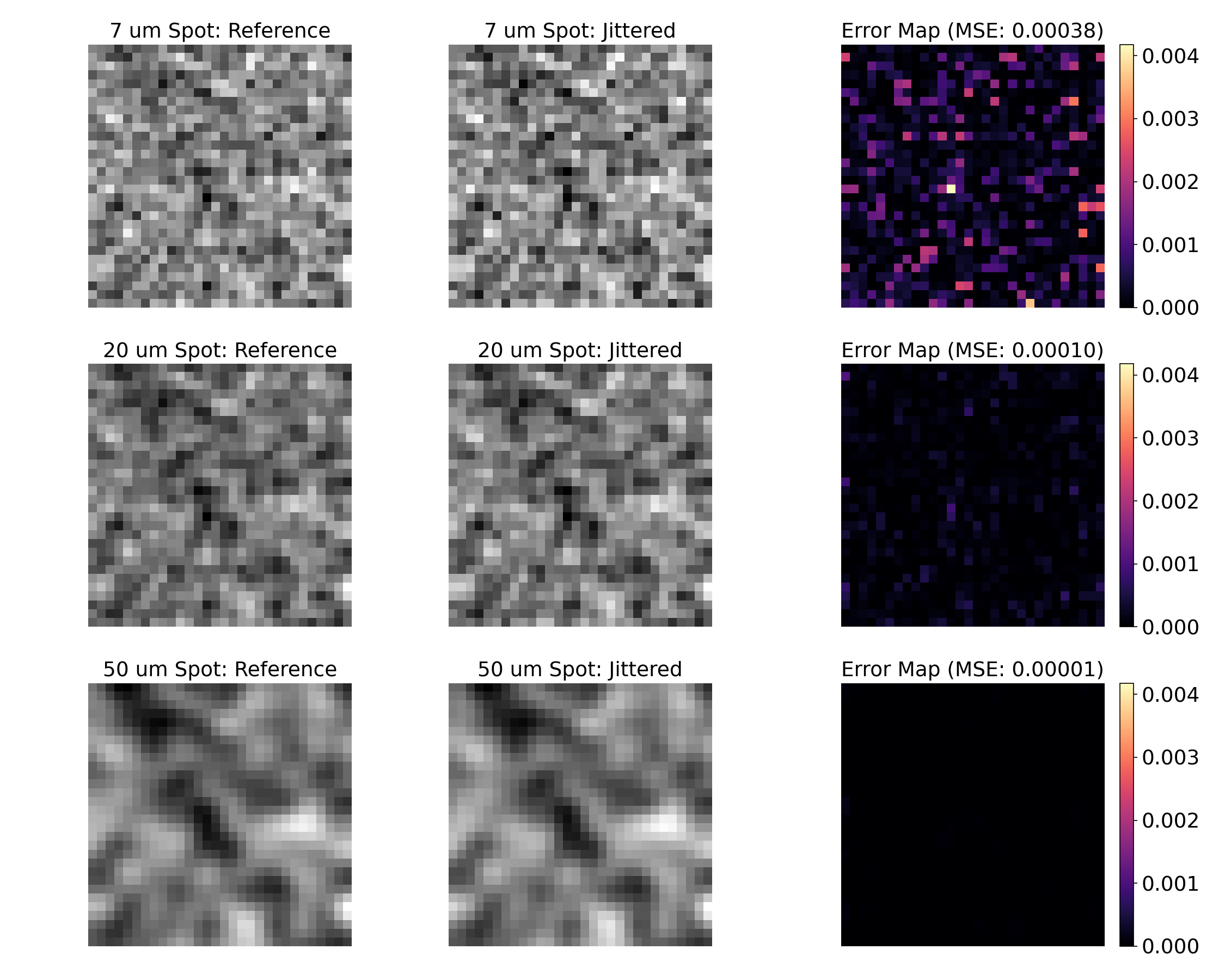} 
    \caption{Representative simulated speckle patterns for various focal spot sizes ($7, 20,$ and $50\ \mu\text{m}$). The columns from left to right display the original speckle patterns, their corresponding jittered counterparts, and the resulting MSE maps, respectively.}
    \label{fig:S1_spatial_images}
\end{figure}

Figure S1 (left column) presents the simulated speckle patterns for focal spot sizes of $7, 20,$ and $50\ \mu\text{m}$, alongside their jittered counterparts (middle column) and corresponding MSE maps (right column). To provide a more comprehensive characterization, the MSE and PCC were quantified as a function of focal spot size, as shown in Fig. S2 and S3, respectively.

\begin{figure}[t]
    \centering
    \includegraphics[width=0.4\linewidth]{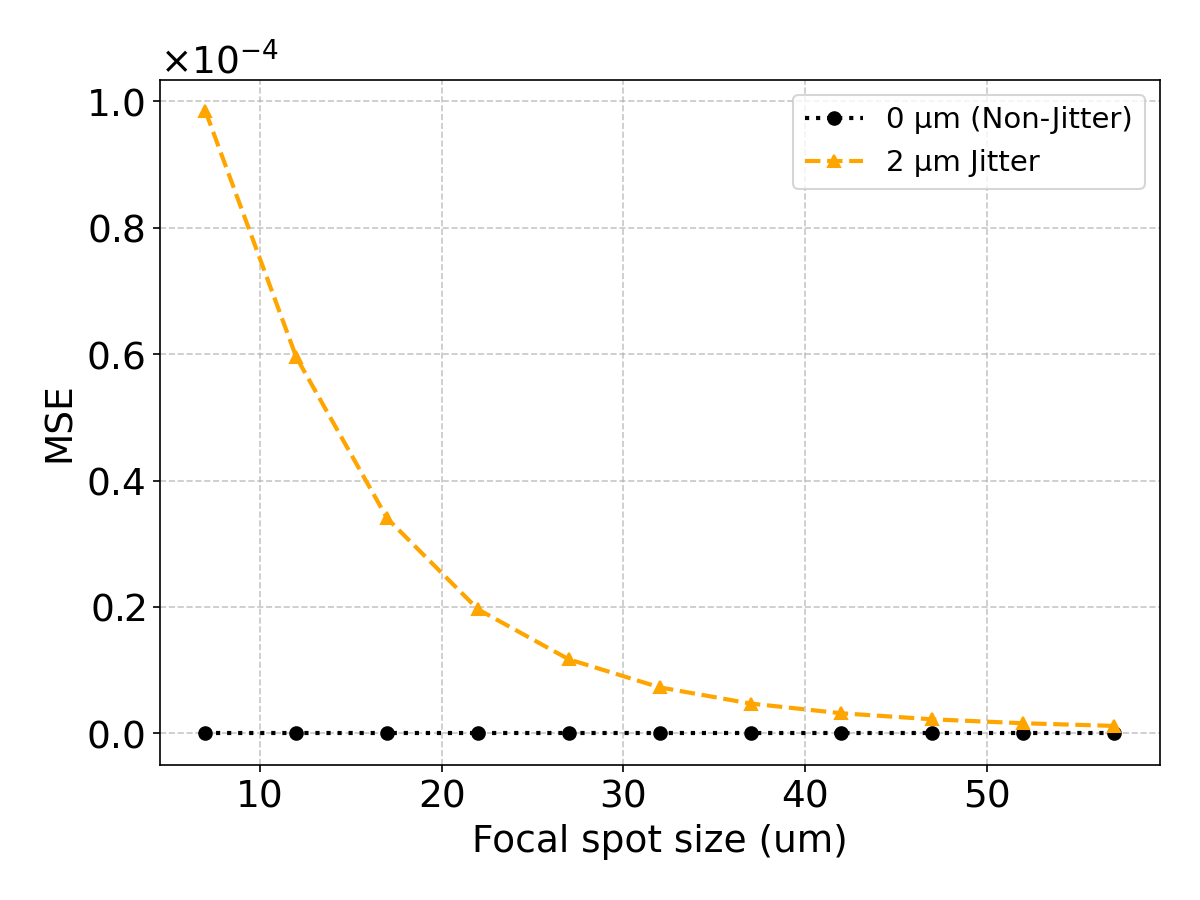} 
    \caption{Mean Squared Error (MSE) as a function of focal spot size.}
    \label{fig:S2_mse_trend}
\end{figure}

\begin{figure}[t]
    \centering
    \includegraphics[width=0.8\linewidth]{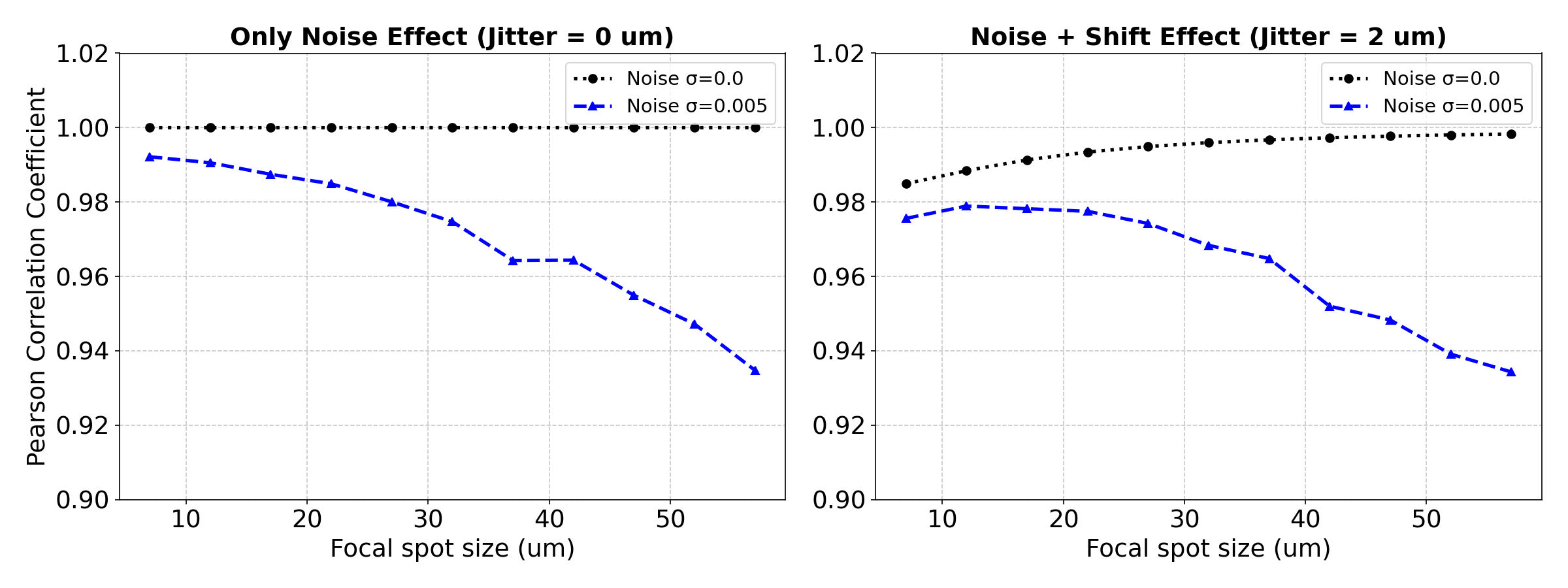} 
    \caption{Pearson Correlation Coefficient (PCC) as a function of focal spot size.}
    \label{fig:S3_pcc_trend}
\end{figure}

The simulation results clearly demonstrate that the $7\ \mu\text{m}$ focal spot yields the highest MSE, with values decreasing as the source size increases. Furthermore, the PCC for the $7\ \mu\text{m}$ source remains comparable to that of the $20\ \mu\text{m}$ source, followed by a decline as the source size increases further. These numerical findings are in excellent agreement with the experimental observations presented in Fig. 4(c) and (d) of the main text, thereby validating the robustness of our retrieval analysis.

\end{document}